\documentclass[aps,prb,reprint,superscriptaddress]{revtex4-2}

\usepackage{graphicx}
\usepackage{dcolumn}
\usepackage{bm}
\usepackage{xcolor}

\begin{document}


\title{UV$_6$Sn$_6$: a new kagome material with unusual \textit{5f} magnetism}

\author{S. M. Thomas}
\affiliation{Los Alamos National Laboratory, Los Alamos, NM 87545}
\author{C. S. Kengle}
\affiliation{Los Alamos National Laboratory, Los Alamos, NM 87545}
\author{W. Simeth}
\affiliation{Los Alamos National Laboratory, Los Alamos, NM 87545}
\author{Chan-young Lim}
\affiliation{Donostia International Physics Center (DIPC), San Sebastián, Spain}
\author{Z. W. Riedel}
\affiliation{Los Alamos National Laboratory, Los Alamos, NM 87545}
\author{K. Allen}
\affiliation{Los Alamos National Laboratory, Los Alamos, NM 87545}
\affiliation{Department of Physics and Astronomy, Rice University, Houston, TX 77005, USA}
\author{A. Schmidt}
\affiliation{Bruker AXS, Madison, Wisconsin 53711,United States}
\author{M. Ruf}
\affiliation{Bruker AXS, Madison, Wisconsin 53711,United States}
\author{Seonggeon Gim}
\affiliation{Department of Physics, Korea Advanced Institute of Science and Technology, Daejeon 34141, Republic of Korea}
\author{J. D. Thompson}
\affiliation{Los Alamos National Laboratory, Los Alamos, NM 87545}
\author{F. Ronning}
\affiliation{Los Alamos National Laboratory, Los Alamos, NM 87545}
\author{A. O. Scheie}
\affiliation{Los Alamos National Laboratory, Los Alamos, NM 87545}
\author{C. Lane}
\affiliation{Los Alamos National Laboratory, Los Alamos, NM 87545}
\author{J. D. Denlinger}
\affiliation{Advanced Light Source, Lawrence Berkeley National Laboratory, Berkeley, CA 94720, USA}
\author{S. Blanco-Canosa}
\affiliation{Donostia International Physics Center (DIPC), San Sebastián, Spain}
\affiliation{IKERBASQUE, Basque Foundation for Science, 48013 Bilbao, Spain}
\author{Jian-Xin Zhu}
\affiliation{Los Alamos National Laboratory, Los Alamos, NM 87545}
\author{E. D. Bauer}
\affiliation{Los Alamos National Laboratory, Los Alamos, NM 87545}
\author{P. F. S. Rosa}
\affiliation{Los Alamos National Laboratory, Los Alamos, NM 87545}

\date{\today}

\begin{abstract}
Materials in the family $R$V$_{6}$Sn$_{6}$ ($R=$ rare earth) provide a unique platform to investigate the interplay between local moments from $R$ layers and nonmagnetic vanadium kagome layers. Yet, the
investigation of actinide members remains scarce. Here we report the synthesis of UV$_{6}$Sn$_{6}$ single crystals through the self-flux technique.
Magnetic susceptibility, specific heat, electrical resistivity, and thermal expansion measurements reveal two uranium-driven antiferromagnetic transitions at $T_{N1}=29$~K and $T_{N2}=24$~K, a complex field-temperature phase diagram, and unusual negative domain wall magnetoresistance.
Specific heat measurements unveil a modest Sommerfeld coefficient of $\gamma = 40$~mJ/mol.K$^{2}$, consistent with angle-resolved photoemission spectroscopy measurements that show a moderate $f$-electron enhancement at the Fermi level ($E_{F}$).
Our experiments support a modest contribution from \textit{5f} flat bands to the density of states at $E_{F}$, whereas our band structure calculations place the vanadium flat bands 0.25~eV above $E_{F}$. Our findings point to a materials opportunity to expand the uranium 166 family with the goal of enhancing correlations by tuning $5f$ and $3d$ flat bands to $E_{F}$.

\end{abstract}

 \maketitle

\section{Introduction}
In quantum materials, many distinct mechanisms can create flat electronic bands, which are realized when the kinetic energy of electrons ($W$) is smaller than the on-site
Coulomb repulsion ($U$), \textit{i.e.}, $U/W > 1$ \cite{Leykam01012018,Regnault2022}. The kinetic energy can be quenched due to band folding (e.g. in moir\'{e} systems\cite{Cao2018}) or quantum interference (e.g. in frustrated
lattices, such as kagome and pyrochlore \cite{Kang2020}), whereas the Coulomb repulsion can be boosted due to reduced dimensionality or strong electronic correlations in local-moment systems (e.g. in $3d$-, $4f$-, and $5f$-based materials \cite{Morosan2012}).
When flat bands occur near or at the Fermi level ($E_{F}$),
sought-after many-body phenomena are known to emerge~\cite{Checkelsky2024}.
Recent examples include the fractional Chern insulating states in moir\'{e} systems~\cite{Xie2021, Zeng2023} and odd-parity superconductivity in $f$-electron UTe$_{2}$~\cite{Ran2019, Aoki_2022}.

Recent developments also highlight the intriguing possibility of combining two distinct flat-band mechanisms in a 
single material. For example, in CuV$_{2}$S$_{4}$, topological flat bands are pinned to the Fermi level due to 
the cooperation between the large on-site Coulomb repulsion from vanadium $3d$ electrons and the quenched kinetic energy due to the pyrochlore lattice geometry~\cite{Huang2024,Rosa2024}.
In layered materials, distinct layer-dependent mechanisms may offer a route towards tunable interlayer coupling.
In particular, the family of layered materials $RM_{6}X_{6}$ ($R$$=$rare earth, $M$$=$transition metal, $X$$=$post transition metal) has recently provided a unique platform to investigate the interplay between highly-localized moments from rare-earth layers and either magnetic or nonmagnetic transition-metal kagome layers, whose electronic structure hosts Dirac crossings, van Hove singularities, and flat bands \cite{Ma2021,Yin2022}.
For example, the magnetic exchange coupling between $3d$ and $4f$ electrons plays an important role in the 
ground state properties of 
$R$Mn$_{6}$Sn$_{6}$ \cite{Riberolles2024,Li2024}, and the uniaxial anisotropy of Tb in 
TbMn$_{6}$Sn$_{6}$ has been argued 
to provide ideal conditions for the opening of a Chern insulating gap in the Dirac dispersion~\cite{Yin2020}.

Given that $5f$ wavefunctions are typically more extended than their $4f$ counterparts, here we investigate the hypothesis that the $d$-$f$ hybridization and resultant flat-band physics may be further tuned, and ultimately enhanced, when the lanthanide layer is replaced by an actinide layer.
To date, however, only two U$M_{6}X_{6}$ members have been reported: 
UCo$_{6}$Ge$_{6}$, which crystallizes in the HfFe$_{6}$Ge$_{6}$-type structure ($P6/mmm$, SG 191)~\cite{Buchholz1981}, and U$_{0.5}$Fe$_{3}$Ge$_{3}$, which crystallizes in the YCo$_{6}$Ge$_{6}$-type structure ($P6/mmm$, SG 191)~\cite{Goncalves1994}. M\"{o}ssbauer measurements of U$_{0.5}$Fe$_{3}$Ge$_{3}$ reveal that the Fe sublattice orders antiferromagnetically (AFM) at $T_{N}=320$~K in a $\mathbf{k} = (0,0,1/2)$ structure. A small ferromagnetic (FM) component at $T_{c} = $220~K is tentatively attributed to the uranium sublattice.

In this paper, we report the synthesis of UV$_{6}$Sn$_{6}$ single crystals through the self-flux technique.
Our magnetic susceptibility, specific heat, electrical resistivity, and thermal expansion measurements in UV$_{6}$Sn$_{6}$ reveal two uranium-driven AFM transitions at $T_{N1}=29$~K and $T_{N2}=24$~K.
Specific heat and angle-resolved photoemission measurements indicate a modest Sommerfeld coefficient and $f$-electron enhancement at $E_{F}$, respectively. UV$_{6}$Sn$_{6}$ displays a complex field-temperature phase diagram and unusual negative domain wall magnetoresistance. Because there are many possible materials combinations within the U$M_{6}X_{6}$ family, our findings provide a route for discovering new members wherein both $5f$ and $3d$ flat bands can be tuned to $E_{F}$.

\begin{figure}[!t]
	\includegraphics[width=\columnwidth]{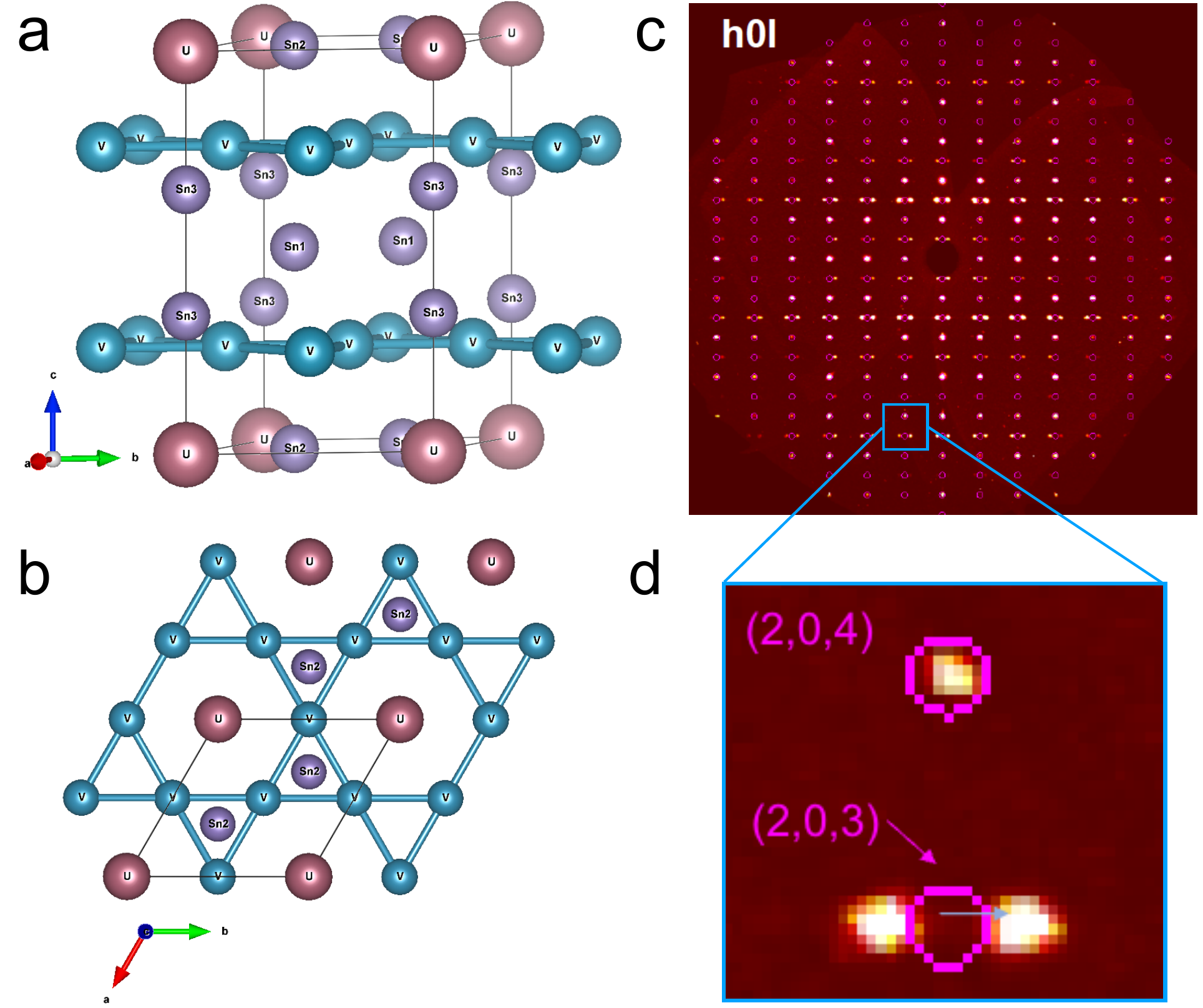}
	\caption{(a) Tentative hexagonal HfFe$_{6}$Ge$_{6}$-type structure of UV$_{6}$Sn$_{6}$ ($P6/mmm$, SG 191) and its (b) $ab$ plane projection, which shows a vanadium kagome layer. (c) An $h0l$ precession image indexed in the hexagonal $P6/mmm$ shows (d) a splitting of the hexagonal peak along $h$ at odd $l$ values, which implies a monoclinic distortion.}
	\label{fig:crystal-structure}
\end{figure}

\section{Experimental Results}

\noindent \textbf{Crystal structure determination} We examine the crystal structure of UV$_{6}$Sn$_{6}$ through laboratory single crystal x-ray diffraction. Because most $R$V$_{6}$Sn$_{6}$ materials in this family are reported to crystallize in the hexagonal HfFe$_{6}$Ge$_{6}$-type structure, our initial unit cell determination attempts considered this parent structure and obtained lattice parameters $a\sim 5.53$ \AA$\,$ and $c \sim 9.17$ \AA, which are very similar to the lattice parameters of LuV$_{6}$Sn$_{6}$ ($a = 5.50$~\AA$\,$ and $c = 9.17$~\AA). As shown in Figure~\ref{fig:crystal-structure}a, this structure consists of two V kagome layers stacked along the crystallographic $c$ axis and separated by triangular layers of U and three inequivalent Sn atoms.

Upon closer inspection, however, a splitting of the hexagonal diffraction peaks is observed in the $h0l$ plane, as shown in the precession images in Figures~\ref{fig:crystal-structure}c,d.
This splitting occurs along $h$ at odd $l$ values and indicates a lowering of the crystal symmetry to monoclinic ($2/m$).
Previous investigations have observed a $2/m$ phase in ErFe$_{6}$Ge$_{3}$Ga$_{3}$ due to an incommensurate structural modulation~\cite{Fredrickson2008}.
The origin of both commensurate and incommensurate modulations in the CoSn-type structure stems from its large interstitial voids that can host electropositive guest atoms, remain empty, or even be half occupied. In many cases, there is an alternation between stuffed and empty cavities along the $c$ axis.
For example, if this alternation is repeated in phase in the ab plane, the HfFe$_{6}$Ge$_{6}$ structure is realized, but if there is a $c$-axis shift by half a unit cell, then the ScFe$_{6}$Ga$_{6}$ structure emerges (\textit{Immm}).
As a result, many types of commensurate and incommensurate superstructures may be created depending on the ordering between filled and empty voids.
As pointed out by Fredrickson \textit{et al.}~\cite{Fredrickson2008}, most 166 structures can be understood as an intergrowth of these two end members.
In the case of ErFe$_{6}$Ge$_{3}$Ga$_{3}$, the monoclinic $2/m$ structure hosts an incommensurate structural wavevector ($\mathbf{q}= 0.092\mathbf{a^{*}} + 0.578\mathbf{b^{*}}$).
Similarly, the monoclinic $P2/m$ structure of UV$_{6}$Sn$_{6}$ exhibits the incommensurate structural wavevector $\mathbf{q}= -0.144\mathbf{a^{*}} + 0.5\mathbf{b^{*}}+ 0.077\mathbf{c^{*}}$.
Here, the q-vector is referenced to the monoclinic unit cell, which has the $b$ and $c$ axes switched compared to the undistorted hexagonal cell.
Additionally, some of the interstitial voids are half occupied.
This implies that the parent compound of UV$_{6}$Sn$_{6}$ is more accurately described by the U$_{0.5}$V$_{3}$Sn$_{3}$-type structure wherein the $c$ axis is halved. As a result, the refined lattice parameters are $a=5.5197(3)$ \AA, $b=4.5857(2)$ \AA, and $c=5.5251(3)$ \AA~\cite{supplemental}.
Further details are provided in the crystallographic information file deposited to the inorganic crystal structure database (CSD 2431850).

In spite of the small monoclinic distortion, UV$_{6}$Sn$_{6}$ crystallizes in hexagonal $ab$ plates with the $c$ axis perpendicular to the plate, akin to most $R$V$_{6}$Sn$_{6}$ materials. For simplicity, we therefore refer to the parent hexagonal crystallographic axes throughout the manuscript.

\begin{figure*}[!ht]
	\includegraphics[width=\textwidth]{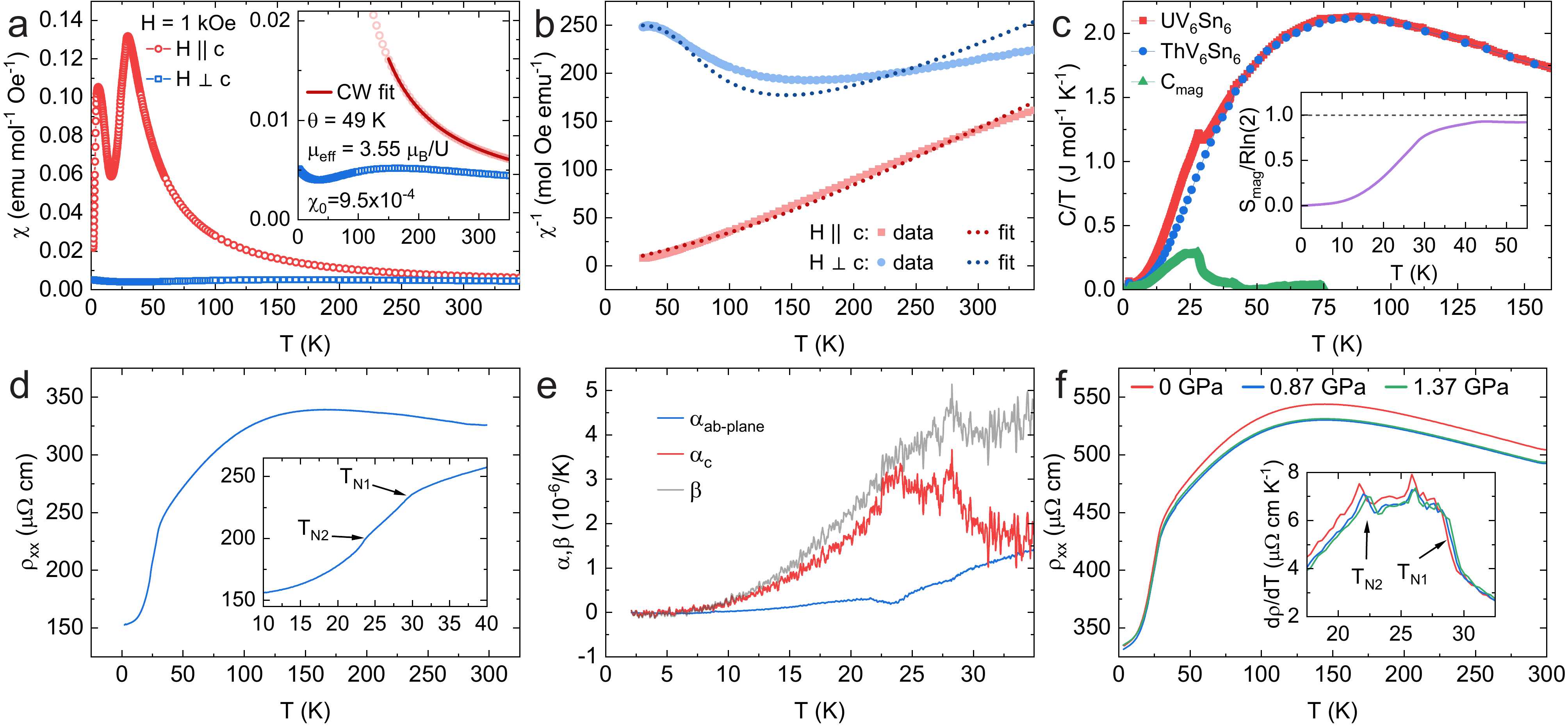}
	\caption{(a)~Temperature-dependent magnetic susceptibility of UV$_{6}$Sn$_{6}$ at a field of 1~kOe applied parallel and perpendicular to the hexagonal $c$ axis.
    (b)~Inverse magnetic susceptibility as a function of temperature at a field of 1~kOe applied parallel and perpendicular to the hexagonal $c$ axis.
    Solid lines are fits to a CEF model.
    (c)~Temperature-dependent specific heat of UV$_{6}$Sn$_{6}$, UV$_{6}$Sn$_{6}$, and the subtracted magnetic contribution. Inset shows the recovered magnetic entropy as a function of temperature.
    (d)~In-plane electrical resistivity as a function of temperature. Inset shows two anomalies associated with magnetic transitions at $T_{N1}=29$~K and $T_{N2}=24$~K.
    (e)~Temperature-dependent thermal expansion of UV$_{6}$Sn$_{6}$.
    (f)~In-plane electrical resistivity of UV$_{6}$Sn$_{6}$ as a function of applied pressure.
    The inset shows the derivative.
    The data in Figs.~(c)--(f) was obtained in zero applied magnetic field.}
	\label{fig:lowfield}
\end{figure*}

\noindent \textbf{Low-field properties} Figure \ref{fig:lowfield}a shows the magnetic susceptibility of UV$_{6}$Sn$_{6}$ as a function of temperature at a field of 1~kOe applied parallel, $\chi_{||}(T)$, and perpendicular, $\chi_{\perp}(T)$, to the hexagonal $c$ axis. Our results reveal a highly anisotropic behavior and suggest that the $c$ axis is the easy axis of magnetization. At high temperatures, a Curie-Weiss (CW) fit to the $c$-axis susceptibility yields an effective moment of 3.55~$\mu_{B}/$U, which is consistent with the Hund's effective moment of either U$^{4+}$ ($\mu_{\text{eff}} = 3.58~\mu_B$) or U$^{3+}$ ($\mu_{\text{eff}} = 3.62~\mu_B$) free ions and implies that the oxidation state of uranium cannot be uniquely determined from CW fits. 

In the simplest approximation using a molecular field in the absence of crystalline electric field (CEF) effects, the positive sign of $\theta_{||} = +49$~K would indicate ferromagnetic interactions between U moments. The highly nonmonotonic behavior of the in-plane susceptibility, however, indicates the presence of CEF effects. In fact, we find that the anisotropy in $\chi(T)$ and the curvature of $\chi_{\perp}(T)$ can be well modeled by a simple hexagonal CEF model (see SI for details \cite{supplemental}) \cite{Stevens1952,Hutchings1964}. Calculations were performed using \textit{PyCrystalField}~\cite{PyCrystalField}, and the results for U$^{3+}$ are shown as solid lines in Fig.~\ref{fig:lowfield}b. All known materials in the $R$V$_{6}$Sn$_{6}$ family host trivalent lanthanides, and in fact the U$^{3+}$ model provides a much better match to the susceptibility compared to the U$^{4+}$ model.

Because of the limitations of fitting susceptibility alone, the extracted CEF parameters are not to be taken at face value.
Nevertheless, after performing several CEF model fits, examining common features of the fitted parameters gives us insight into the single-ion physics in UV$_{6}$Sn$_{6}$. 
Examining the fitted eigenstates of the U$^{3+}$ models, we find that all models (regardless of how many Stevens parameters were varied) converge to $B_2^0 < 0$, which is consistent with an easy $c$ axis, and a ground state doublet dominated by $|\pm 9/2 \rangle$ (see SI for details \cite{supplemental}). The models that most closely reproduce the data, which include $B_4^0$ and higher, yield a first excited CEF level between 18~meV and 25~meV, consistent with well-isolated easy-axis spins. 

These CEF models are purely single-ion Hamiltonians, which neglect magnetic exchange or hybridization with itinerant electrons.
As a result, the CEF model does not match the experimental susceptibility at low temperatures.
A clear downturn is observed in $\chi_{||}(T)$ at $T_{N1}=29$~K, typical of an antiferromagnetic (AFM) transition when magnetic field is applied along the easy axis.
A small change in slope is observed at $T_{N2}=24$~K, which suggests a change in the magnetic structure.
Finally, we note the presence of a clear third anomaly at $\sim 4$~K.
As we will see below, this anomaly is suppressed with magnetic fields and does not display a corresponding specific heat anomaly.
We therefore conclude that it is  caused by residual superconducting V$_{3}$Sn inclusions~\cite{Okubo1968}.

Figure \ref{fig:lowfield}c shows the temperature dependence of the specific heat, C/T, for UV$_{6}$Sn$_{6}$ and its nonmagnetic analog, ThV$_{6}$Sn$_{6}$.
A clear lambda-type anomaly is observed at $T_{N1}=29$~K, in agreement with magnetic susceptibility results.
In contrast, only a small inflection is observed at $T_{N2}=24$~K, which indicates a small change in the magnetic configuration.
No transitions are observed in ThV$_{6}$Sn$_{6}$, as expected from nonmagnetic vanadium layers.
The magnetic component of the specific heat (green triangles), C$_{\mathrm{mag}}$/T, is obtained by subtracting the nonmagnetic contribution from the total specific heat of UV$_{6}$Sn$_{6}$.
The entropy recovered at $T_{N1}=29$~K, $S(T_{N1})$, is about 80~\% of $R$ln2, whereas the full $R$ln2 entropy is recovered around 45~K.
A modest reduction in $S(T_{N1})$ is typically an indication of either magnetic frustration or Kondo screening.
A small electronic specific heat coefficient ($\gamma = 40$ mJ~K$^{-2}$~mol$^{-1}$), determined by fitting the linear portion of $C_p/T$ versus $T^2$ at low temperatures, indicates that Kondo screening is weak in UV$_{6}$Sn$_{6}$.
The modest entropy recovered at $T_{N1}$ therefore likely arises from short-range correlations above $T_{N1}$.

The temperature-dependent in-plane resistivity, $\rho_{xx}(T)$, is shown in Figure \ref{fig:lowfield}d.
At high temperatures, $\rho_{xx}(T)$ displays a slight increase on cooling, consistent with incoherent Kondo scattering.
Below 150~K, $\rho_{xx}(T)$ starts to decrease on cooling, and the broad anomaly centered around 170~K is typical of depopulation of excited CEF levels.
Two clear kinks are observed at $T_{N1}=29$~K and $T_{N2}=24$~K, in agreement with specific heat measurements.
Below 10~K, $\rho_{xx}(T)$ can be well described by a Fermi-liquid (FL) term, $\rho(T) = \rho_0 + AT^2$, with $\rho_0 = 152 ~\mu\Omega \text{cm}$ and $A = 0.036~\mu\Omega\text{cmK}^{-2}$. 
By combining the FL coefficient and the Sommerfeld coefficient, we extract the Kadowaki-Woods (KW) ratio $\frac{A}{\gamma^2} \approx 2.2\times 10^{-5}\mu\Omega\text{cm} (\frac{\text{mol K}}{\text{mJ}})^2$, which is consistent with $f$-electron materials  \cite{Kadowaki}.

Figure \ref{fig:lowfield}e shows the temperature-dependent thermal expansion performed using a capacitance dilatometer described in Ref.~\cite{Schmiedeshoff2006a}. The linear expansion coefficient along the $c$ axis, $\alpha_{c}(T)$, displays a pronounced positive jump at $T_{N1}$, consistent with a second-order phase transition, followed by a smaller expansion at $T_{N2}$. In contrast, $\alpha_{ab}(T)$ shows a much more modest response: a small negative jump at $T_{N2}$ and an almost undetectable negative jump at $T_{N1}$. As a result, the volume thermal expansion, $\beta$ is dominated by the positive $c$-axis contribution, a finding that is also consistent with spins pointing along the \textit{c} axis. The pressure ($P$) dependence of a second-order phase transition can be estimated through the Ehrenfest relation, $dT_{N1}/dP = V_{m}\Delta \beta /(\Delta C_{p}/T_{N1})$, where $V_{m}$ is the molar volume, $\Delta\beta$ is the jump in volume thermal expansion at $T_{N1}$, and $\Delta C_{p}$ is the jump in heat capacity. Because $C_{p}$ is always positive, the sign of the pressure dependence is determined by the sign of the jump in volume thermal expansion. The positive jump in UV$_{6}$Sn$_{6}$ yields an increase in $T_{N1}$ with pressure, $dT_{N1}/dP \sim 0.4$~K/GPa. In fact, our pressure-dependent electrical resistivity measurements, shown in Figure \ref{fig:lowfield}f, reveal that $dT_{N1}/dP = 0.33$~K/GPa, in good agreement with the Ehrenfest relation.

\begin{figure}[!t]
	\includegraphics[width=\columnwidth]{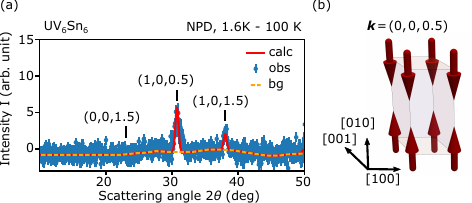}
    \caption{
        (a) Magnetic neutron powder diffraction (NPD) data.
        Diffraction intensity (blue data points) was recorded with neutrons of wavelength $\lambda=2.45\mathrm{\AA}$ and temperature subtracted ($T=1.6-100$K).
        The observed intensity (obs) was fit with an A-type antiferromagnetic structure of wave-vector $\bm{k}=(0,0,1/2)$ and moments pointing along the c-axis (red line, calc).
        For the refinement, a scattering-angle dependent background defined manually was considered (orange interrupted line, bg).
        Momentum transfers of several Bragg peaks are indicated by vertical lines. 
        Reciprocal coordinates are defined according to a hexagonal lattice with parameters $a=5.52\,\mathrm{\AA}$ and $c=9.17\,\mathrm{\AA}$ .
        (b) Spin texture as considered for the refinement in (a).  
    }
	\label{fig:mag-structure}
\end{figure}

\noindent \textbf{Magnetic structure determination:} Our magnetic susceptibility, specific heat, and thermal expansion measurements all indicate an AFM structure with moments pointing along the $c$ axis.
To solve the structure, we performed powder neutron diffraction on UV$_{6}$Sn$_{6}$ both in the paramagnetic state ($T = 100$~K) and at low temperatures ($T = 2$~K)~\cite{SINQ}.
The temperature-subtracted data, shown in Figure~\ref{fig:mag-structure}a, reveal magnetic peaks that can be indexed by an A-type AFM structure with wavevector $\mathbf{k} = (0,0,1/2)$ and uranium moments pointing along the $c$ axis. As shown in Figure~\ref{fig:mag-structure}b, uranium moments align ferromagnetically in the $ab$ plane and antiferromagnetically along the $c$ axis. Typically, A-type magnetic structures require at least two exchange interactions, and our results are consistent with the presence of an in-plane ferromagnetic interaction competing with an antiferromagnetic interaction along the $c$ axis.

\begin{figure*}[!ht]
	\includegraphics[width=1\textwidth]{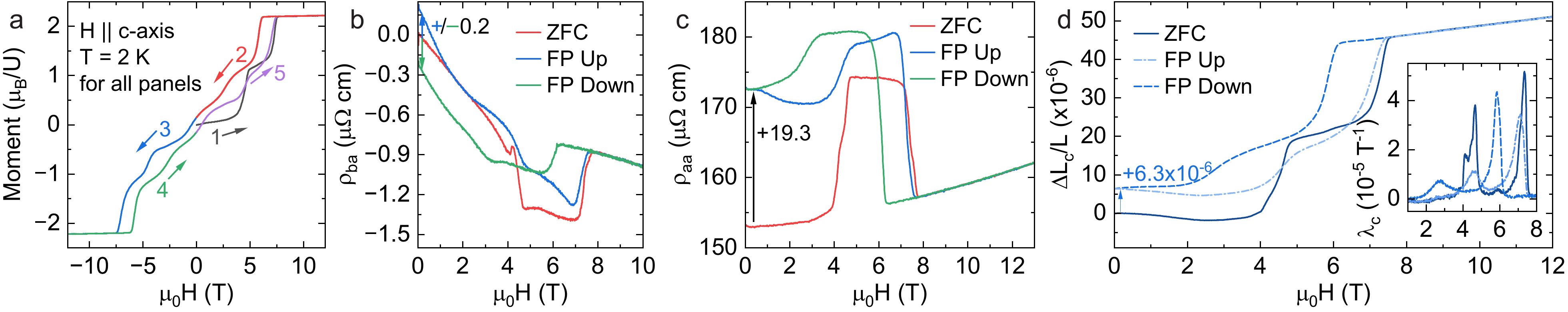}
	\caption{
        Field-dependent data collected at 2~K:
        (a) magnetization,
        (b) Hall resistivity ($\rho_{ba}$),
        (c) longitudinal resistivity ($\rho_{aa}$),
        (d) change in length along the $c$ axis divided by the length after cooling in zero field.
        The inset of shows $\lambda_c=(dL_c/dH)/L_{c,0}$.
        For all measurements the field is applied along the $c$ axis.
        ZFC data was collected by cooling the sample in zero field and then ramping to the indicated field.
        ``FP Up'' describes starting in the field-polarized state, ramping back to zero field, and then ramping to the indicated field.
        ``FP Down' describes starting in the field-polarized state and then ramping to the indicated field.
        Resistivity data was symmetrized or anti-symmetrized by performing the same procedure but with field of opposite polarity.
        The number indicates the change in quantity at zero field between ZFC and FP conditions.
    }
	\label{fig:2kdata}
\end{figure*}

\noindent \textbf{Magnetic field-dependent properties:} Despite the simple A-type magnetic structure at zero field, UV$_{6}$Sn$_{6}$ displays unusual magnetic properties. First, remanent magnetization ($M_{\mathrm{rem}}$) is observed in magnetization loops at low temperatures, as shown in Figure~\ref{fig:2kdata}a, a result that is at odds with a magnetic structure that has zero net magnetization. We note that $M_{\mathrm{rem}} =0.1575\mu_{B}$ translates to $\sim 1/13$ of the total saturation magnetization, $M_{s}=2.167\mu_{B}$. At first sight, this finding suggests that there is 1 ferromagnetic plane out of $13$ planes that remains uncompensated, which is consistent with a modulated structure with wavevector $\mathbf{k} = (0,0,7/13)$ at zero field corresponding to the following $c$-axis stacking: 
(\textcolor{red}{$\uparrow\uparrow$}\textcolor{blue}{$\downarrow\downarrow$}\textcolor{red}{$\uparrow\uparrow$}\textcolor{blue}{$\downarrow\downarrow$}\textcolor{red}{$\uparrow\uparrow$}\textcolor{blue}{$\downarrow\downarrow$}$\uparrow$). Though the inferred value $k = 7/13 = 1/2 + \delta$ ($\delta = 0.038$) is very close to $k=1/2$, the instrumental resolution of our neutron diffraction measurements allows us to rule out such a modulated structure. Another possibility is history dependence of the magnetic properties. For instance, the zero-field magnetic ground state may change after field ramping through field-induced transitions, as observed previously in Fe$_{1-x}$Co$_{x}$Si \cite{Bauer2016}. However, $M_{\mathrm{rem}}$ is observed even if field sweeps do not cross phase boundaries (Fig.~S1 in the SI). As a result, we conclude that there must be a small uniform magnetization component along the $c$ axis on top of the $\mathbf{k} = (0,0,1/2)$ structure, consistent with a ferrimagnetic ground state. We note that the absence of a net moment in zero-field cooled conditions is likely due to the presence of domains at zero field, as discussed in the SI. 

At higher fields, our magnetization data display a plateau at $H_{p} \sim 4.5$~T, consistent with a field-induced transition to a different ordered phase (AFM3).
The plateau magnetization is roughly $M_{s}/2$, which points to a wavevector $\mathbf{k} = (0,0,3/4)$ and $c$-axis stacking (\textcolor{red}{$\uparrow\uparrow\uparrow$}\textcolor{blue}{$\downarrow$}).
At $H_{s}=7.5$~T, UV$_{6}$Sn$_{6}$ enters a fully-polarized state. The large hysteresis in magnetization loops suggests highly first-order field-induced transitions and domain physics. Our results resemble those in `Devil's staircase' CeSb wherein all magnetic phases consist of square-wave structures characterized by a wavevector $\mathbf{k} = (0,0,k)$ commensurate with the crystal lattice~\cite{Rossat1977}.
This behavior can be modeled by a spin-1/2 axial next-nearest-neighbor Ising (ANNNI) model with in-plane ferromagnetic coupling  ($J_{0}$) and competing ferromagnetic and antiferromagnetic interactions between nearest ($J_{1}$) and next-nearest ($J_{2}$) layers along the $c$ axis~\cite{selke1988}.

We now turn to field-dependent electrical transport properties.
Figure~\ref{fig:2kdata}b shows Hall resistivity as a function of fields applied along the $c$ axis.
Akin to the magnetization loops, Hall data reveal anomalies at $H_{p}$ and $H_{s}$ and pronounced hysteresis.
We note that it was not possible to fit the Hall resistivity assuming field independent anomalous and ordinary contributions, indicating that the Fermi surface likely changes as function of magnetic field~(See Fig. S2 in the SI \cite{supplemental}). The finite sign-changing Hall intercept in upsweeps or downsweeps confirms the presence of an anomalous component due to the uncompensated ferromagnetic contribution discussed above.

Figure~\ref{fig:2kdata}c shows the in-plane electrical resistivity, $\rho_{aa}$, as a function of fields applied along the $c$ axis.
As expected, $\rho_{aa}$(H) also reveals anomalies at $H_{p}$ and $H_{s}$ as well as pronounced hysteresis, likely due to domain physics.
Contrary to expectations, however, the zero-field intercept of $\rho_{aa}$ after field sweeps (FP) is significantly higher compared to its value in zero-field-cooled (ZFC) conditions.
Typically, domains are expected to cause an increase in electrical resistance due to scattering of electrons by the domain walls, which act as effective magnetic barriers.
A ZFC state, which displays a higher density of domains, would therefore imply a higher resistance state, which is at odds with our observations. 
In ferromagnets, semiclassical calculations reveal that domain wall resistance may be negative in special cases when the scattering lifetimes are spin dependent \cite{Gorkom1999}.
In antiferromagnetic metals, recent magnetoresistance calculations show that the domain-wall magnetoresistance (DWMR) can be negative in the diffusive regime, wherein the domain-wall width is larger than the mean free path, depending on the domain-wall orientation and spin structure \cite{Zheng2020}, which is a sensible scenario in UV$_{6}$Sn$_{6}$.

\begin{figure}[b]
	\includegraphics[width=0.9\columnwidth]{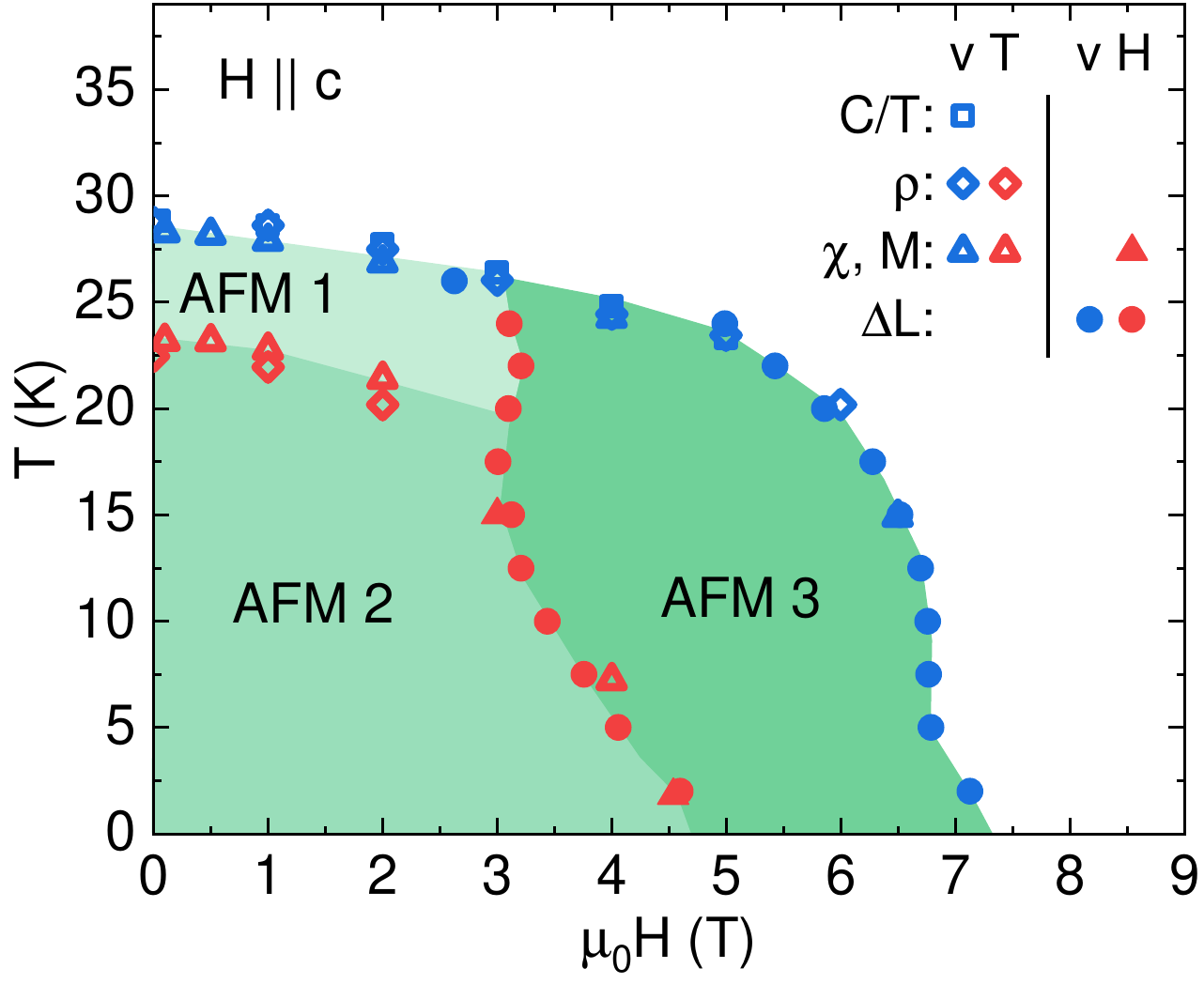}
	\caption{
        Magnetic field versus temperature phase diagram for field parallel to the $c$ axis.
        Open symbols were collected by cooling from above T$_N$ to base temperature.
        Closed symbols were collected by first cooling in zero field and then performing a field ramp toward larger field.
    }
	\label{fig:phaseD}
\end{figure}

\begin{figure*}[!ht]
	\includegraphics[width=0.8\textwidth]{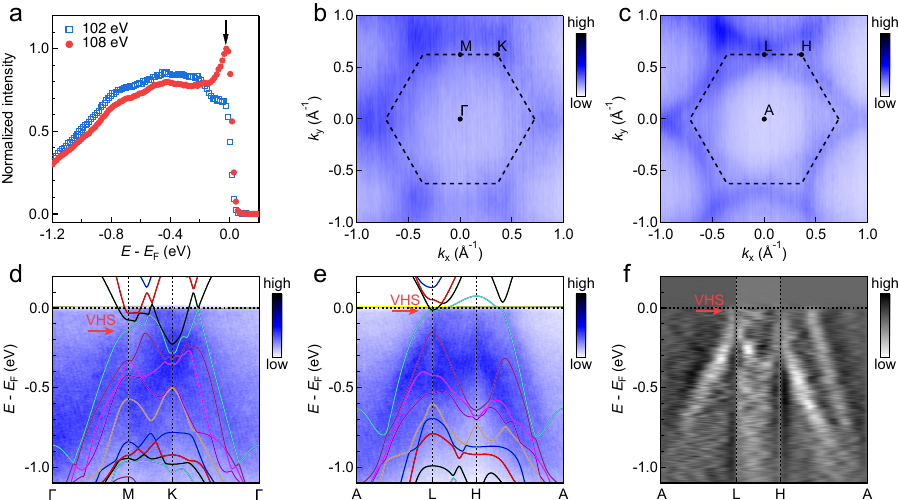}
	\caption{Electronic structure of UV$_{6}$Sn$_{6}$ measured using ARPES. (a) Momentum-integrated EDCs of K-$\Gamma$-K dispersions measured under on-resonance (108 eV, red filled circles) and off-resonance (102 eV, blue hollow squares) conditions. EDCs are normalized with respect to the maximum point of 108 eV EDC (b,c) Fermi surfaces at $\Gamma$ (b, 102 eV) and $A$ (c, 88 eV) planes. Black dashed lines show Brillouin zone boundaries. (d,e) Valence band dispersions along the $\Gamma$-M-K-$\Gamma$ (d, 102 eV) and A-L-H-A (e, 88 eV) directions. The DFT-calculated bands in the paramagnetic phase are overlaid on the ARPES data. (f) 2nd derivative of (e).}
	\label{fig:ARPES}
\end{figure*}

Further information about domain formation in UV$_{6}$Sn$_{6}$ is obtained from longitudinal magnetostriction measurements along the $c$ axis, $\Delta L_{c}/L$, shown in Figure~\ref{fig:2kdata}d.
The zero-field intercept of $\Delta L_{c}/L$ after field sweeps (FP) is higher compared to its value in zero-field-cooled (ZFC) conditions, which implies that domains yield a shorter sample length.
Considering that the spin structure of UV$_{6}$Sn$_{6}$ is aligned along the $c$ axis, this length difference is rather unexpected because domains with antiparallel magnetization do not differ in their magnetoelastic strains.
We note, however, that spins rotate away from the $c$ direction within domain walls (either Bloch or Ne\'{e}l).
As a result, we infer that the domain wall volume is large enough in ZFC conditions to contribute an in-plane moment to the magnetoelastic strain, which shortens the sample length.

Our combined magnetic susceptibility, specific heat, electrical transport, and thermal expansion data give rise to the temperature-field phase diagram shown in Figure~\ref{fig:phaseD} (see Fig.~S3 for additional field-dependent data \cite{supplemental}). Magnetization and neutron diffraction data point to a ground state AFM2 configuration with a wavevector $\mathbf{k} = (0,0,1/2)$, whereas magnetization data suggest that AFM3 has a wavevector of $\mathbf{k} = (0,0,3/4)$. Our thermodynamic data and CEF fits further indicate that the properties of UV$_{6}$Sn$_{6}$ are dominated by U$^{3+}$ $5f$ local moments with weak Kondo renormalization.

\section{Band Structure Calculations and Angle-Resolved Photoemission}
To address whether flat bands from vanadium kagome layers are close to the Fermi level and whether there is measurable $5f$-weight at $E_{F}$, we perform angle-resolved photoemission spectroscopy (ARPES) measurements and density functional theory (DFT) calculations.
The Fermi maps were taken on the Sn termination as verified by x-ray photoelectron spectroscopy (XPS) scans (see Fig.~S4 for details \cite{supplemental}) \cite{Lee2024}.
Fig.~\ref{fig:ARPES}a shows the momentum-integrated energy distribution curves (EDCs) taken at $T=40$~K from K-$\Gamma$-K dispersions with both on-resonance ($h\nu= 108$~eV, red circles) and off-resonance ($h\nu= 102$~eV, blue squares) conditions for the uranium $M$ edge.
A small but clear resonance enhancement is detected near the Fermi level, which indicates the presence of hybridization between $5f$ and conduction electrons but with a smaller strength compared to prototypical heavy-fermion materials such as CeCoIn$_{5}$ \cite{Chen2017,Jang2020}. We note that the surface termination (Sn) and the small $5f$ relative cross section with respect to vanadium $d$ bands also likely contribute to the weak enhancement.

The Fermi surface exhibits a dispersionless band along the $c$ direction, which demonstrates its marked two-dimensional character (see Fig~S5 for details \cite{supplemental}) and enables the location of the high-symmetry points $k_{z} = 0$ ($\Gamma$) and at $k_{z} = \pi/c$ ($A$) points at 102 and 88~eV, respectively.
The Fermi contours shown in Figures~\ref{fig:ARPES}b-c reveal the expected six-fold symmetry characteristic of the parent kagome structure as well as triangular pockets at K and H, which are most visible at $k_{z} = \pi/c$ (Figure~\ref{fig:ARPES}c).
Figures~\ref{fig:ARPES}d-e display the valence band (VB) spectra along the high-symmetry directions $\Gamma$-M-K-$\Gamma$ ($k_{z} = 0$) and A-L-H-A ($k_{z} = \pi/c$), respectively. The electronic band structure in the $k_{z} = 0$ plane, shown in Figures~\ref{fig:ARPES}d, is dominated by multiple bands that form a large electron pocket centered at $\Gamma$ and smaller pockets around $K$ and $M$, in reasonable agreement with the overlaid DFT calculations.
The spectra in the L-H-A-L plane more clearly reveal two linearly dispersing bands from which we extract the electron velocities $v_{F} = 3.19 \times 10^{5}$~m/s (inner pocket) and $v_{F} = 2.34 \times 10^{5}$~m/s (outer pocket). These Fermi velocities values are slightly larger those for ScV$_{6}$Sn$_{6}$, $v_{F} = 1.5-2 \times 10^{5}$~m/s \cite{Yi2024}, which  supports the weak presence of heavy $5f$ electrons at $E_{F}$.
There is somewhat poor agreement between the ARPES data and the DFT-calculated band structure at the H point near -0.25~eV.
This signal is unlikely to originate from a surface state due to its strong ARPES intensity and is also unrelated to magnetic ordering, as the measurement was conducted at 40 K, which is above the ordering temperatures.
Instead, it may be related to the fact that the DFT calculations were performed using the HfFe$_6$Ge$_6$-type structure and do not take into account the distortion that lowers the symmetry to $2/m$.

As expected from the kagome structure, the VB spectra around the M (L) points reveal van Hove singularities (vHS) very close to $E_{F}$, which become more pronounced in their second derivative along the A-L-H-A direction, shown in Figure~\ref{fig:ARPES}f. 
The spectra along $\Gamma$-M-K-$\Gamma$ and A-L-H-A, however, do not display indications of flat bands below the Fermi level, either from vanadium or uranium bands, in agreement with DFT calculations. Our results hence support to the prominent role of U in the magnetic transitions. 

\begin{figure}[b]
	\includegraphics[width=\columnwidth]{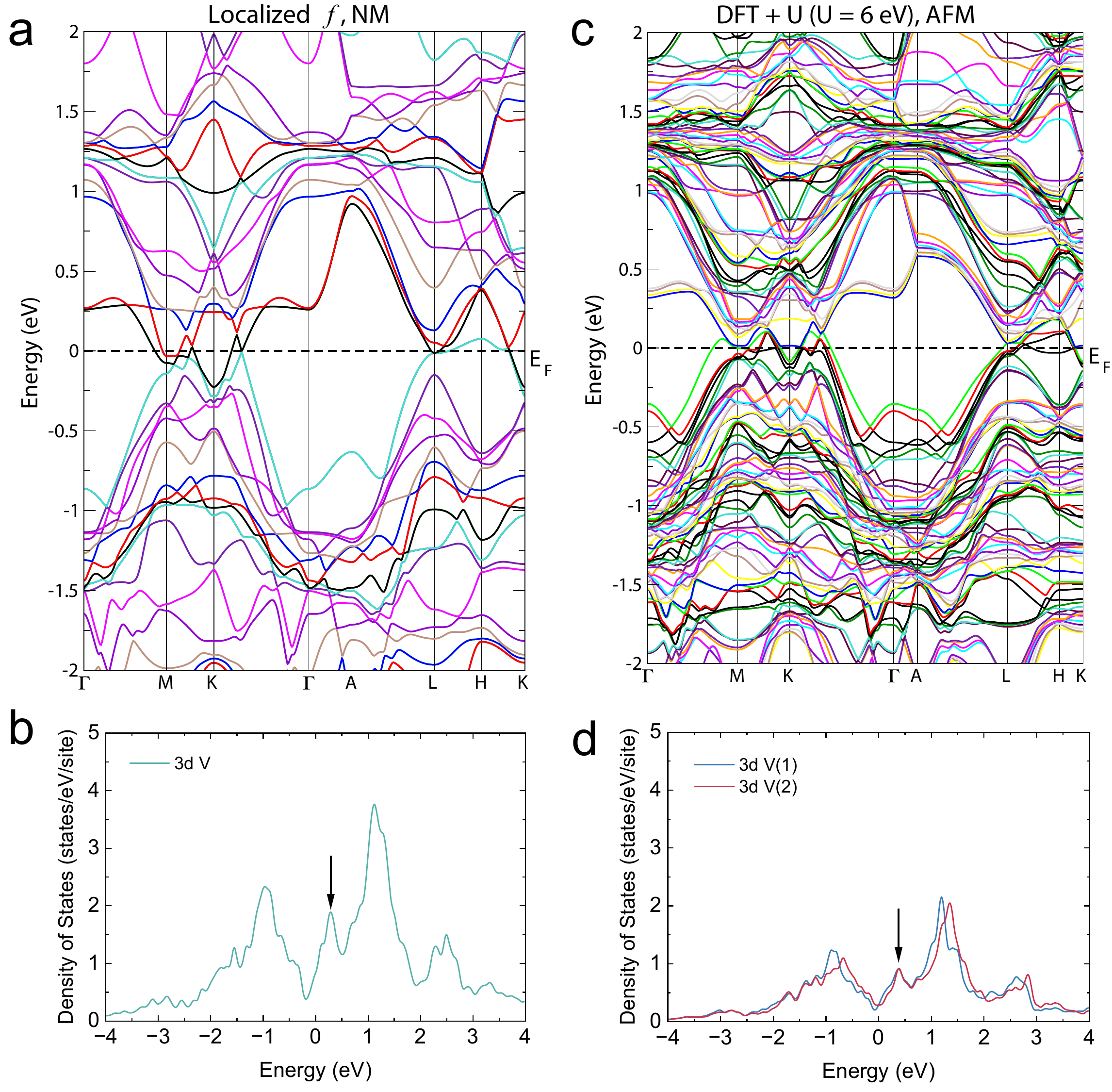}
	\caption{
        a) DFT band structure calculation of UV$_{6}$Sn$_{6}$ in the nonmagnetic state with localized $f$ electrons and b) the associated V $3d$ partial density of states. c) DFT+U ($U=6$~eV) band structure calculation of UV$_{6}$Sn$_{6}$ in the antiferromagnetic state and d) the associated V $3d$ partial density of states.
        Spin-orbit coupling is included in both calculations.
    }
	\label{fig:DFT}
\end{figure}

Our DFT calculations utilized the generalized gradient approximation (GGA) and the Perdew-Burke-Ernzerhof (PBE) exchange correlational function with the WIEN2k package to obtain the electronic band structure of UV$_{6}$Sn$_{6}$ \cite{SCHWARZ2003,Perdew1996}. For simplicity, we used the HfFe$_{6}$Ge$_{6}$-type structure with lattice parameters $a\sim 5.53$ \AA$\,$ and $c \sim 9.17$ \AA.
Because experimental data indicate that uranium $5f$ electrons are localized, DFT calculations investigate two scenarios: Figure~\ref{fig:DFT}a shows the band structure of UV$_{6}$Sn$_{6}$ in the paramagnetic state assuming that three uranium $5f$ electrons are localized in the core, whereas Figure~\ref{fig:DFT}b shows the DFT+U band structure calculation in the AFM state (\textcolor{red}{$\uparrow\uparrow$}\textcolor{blue}{$\downarrow\downarrow$}) with $U=6$~eV.
In the first calculation, two dispersive vanadium bands cross the Fermi level in the $k_{z}=0$ plane ($\Gamma-M-K-\Gamma$ path) and create electron pockets, in agreement with electron-dominated Hall effect data in the paramagnetic state. In addition, a van Hove singularity is observed at the $M$ point below the Fermi level.

Flat bands are identified in the $k_{z}=0$ plane 0.25~eV above $E_{F}$.
These flat bands arise from the vanadium kagome layers and give rise to a peak in the vanadium $3d$ partial density of states at 0.25~eV, as shown in Figure~\ref{fig:DFT}c (black arrow). In the DFT+U calculation, shown in Figure~\ref{fig:DFT}b, two additional bands cross $E_{F}$ in the $k_{z}=0$ plane, and the flat bands from vanadium kagome layers shift to slightly higher energies.
The density of states (DOS) as a function of energy, shown in Figure~\ref{fig:DFT}d, confirms that the main contribution at the Fermi level comes from V bands, and that the flat bands are located at 0.35~eV above the Fermi level (black arrow). We note that our calculations are qualitatively similar to those recently reported for ThV$_{6}$Sn$_{6}$ \cite{Xiao2024}; however, the total DOS at $E_{F}$ in our localized calculation, 7~states/eV/site, is roughly 40\% larger than the total DOS in ThV$_{6}$Sn$_{6}$. This difference could account for the discrepancy between the calculated and experimental Sommerfeld coefficients in ThV$_{6}$Sn$_{6}$.

In summary, we report the synthesis of UV$_{6}$Sn$_{6}$ single crystals through the self-flux technique.
Our zero-field magnetic susceptibility, specific heat, electrical resistivity, and thermal expansion measurements in UV$_{6}$Sn$_{6}$ reveal two uranium-driven antiferromagnetic transitions at $T_{N1}=29$~K and $T_{N2}=24$~K. Magnetic powder diffraction data point to an antiferromagnetic ground state with wavevector $\mathbf{k} = (0,0,1/2)$ and uranium moments pointing along the $c$ axis. As a function of applied magnetic fields along the \textit{c} axis, we observe unusual behavior, including remanent magnetization and negative domain wall resistance. Specific heat measurements unveil a modest Sommerfeld coefficient of $\gamma = 40$~mJ/mol.K$^{2}$, in agreement with a moderate $f$-weight enhancement observed in angle-resolved photoemission spectroscopy measurements. Our experiments support a modest contribution from \textit{5f} flat bands at $E_{F}$, whereas our band structure calculations place the vanadium flat bands 0.25~eV above $E_{F}$. Our findings point to a materials opportunity to expand the U$M_{6}X_{6}$ family with the goal of tuning both $5f$ and $3d$ flat bands close to the Fermi level.

Note: During the preparation of this manuscript, we became aware of an independent experimental study of UV$_6$Sn$_6$, which also revealed a (0, 0, 1/2) ground state magnetic structure with moments along the $c$ axis~\cite{raymond2024}.

\begin{acknowledgments}

Work at Los Alamos National Laboratory was performed under the auspices of the U.S. Department of Energy, Office of Basic Energy Sciences, Division of Materials Science and Engineering.
CSK, WS, and ZR acknowledge support from the Laboratory Directed Research and Development program. Scanning electron microscope and energy dispersive x-ray measurements were performed at the Electron Microscopy Lab and supported by the Center for Integrated Nanotechnologies, an Office of Science User Facility operated for the U.S. Department of Energy Office of Science.
The electronic structure calculations were supported in part by the Center for Integrated Nanotechnologies, a DOE BES user facility, in partnership with the LANL Institutional Computing Program for computational resources. Additional computations were performed at the National Energy Research Scientific Computing Center (NERSC), a U.S. Department of Energy Office of Science User Facility located at Lawrence Berkeley National Laboratory, operated under Contract No. DE-AC02-05CH11231 using NERSC award ERCAP0028014.
S. Blanco-Canosa thanks financial support from the MINECO of Spain through the project PID2021-122609NB-C21. CY Lim acknowledges the European Research Council (ERC) under the European Union’s Horizon 2020 research and innovation program (grant agreement no. 101020833.
This research used resources of the Advanced Light Source, which is a DOE Office of Science User Facility under contract no. DE-AC02-05CH11231.
This work is partly based on experiments performed at the Swiss spallation neutron source SINQ, Paul Scherrer Institute, Villigen, Switzerland.
The authors thank D. Sheptyakov for support with the neutron powder diffraction experiments.

\end{acknowledgments}

\bibliography{lib}

\end{document}


\title{Supplemental Information for `UV$_6$Sn$_6$: a new kagome material with unusual \textit{5f} magnetism'} 

\author{S. M. Thomas}
\affiliation{Los Alamos National Laboratory, Los Alamos, NM 87545}
\author{C. S. Kengle}
\affiliation{Los Alamos National Laboratory, Los Alamos, NM 87545}
\author{W. Simeth}
\affiliation{Los Alamos National Laboratory, Los Alamos, NM 87545}
\author{Chan-young Lim}
\affiliation{Donostia International Physics Center (DIPC), San Sebastián, Spain}
\author{Z. W. Riedel}
\affiliation{Los Alamos National Laboratory, Los Alamos, NM 87545}
\author{K. Allen}
\affiliation{Los Alamos National Laboratory, Los Alamos, NM 87545}
\affiliation{Department of Physics and Astronomy, Rice University, Houston, TX 77005, USA}
\author{A. Schmidt}
\affiliation{Bruker AXS, Madison, Wisconsin 53711, United States}
\author{M. Ruf}
\affiliation{Bruker AXS, Madison, Wisconsin 53711, United States}
\author{Seonggeon Gim}
\affiliation{Department of Physics, Korea Advanced Institute of Science and Technology, Daejeon 34141, Republic of Korea}
\author{J. D. Thompson}
\affiliation{Los Alamos National Laboratory, Los Alamos, NM 87545}
\author{F. Ronning}
\affiliation{Los Alamos National Laboratory, Los Alamos, NM 87545}
\author{A. O. Scheie}
\affiliation{Los Alamos National Laboratory, Los Alamos, NM 87545}
\author{C. Lane}
\affiliation{Los Alamos National Laboratory, Los Alamos, NM 87545}
\author{J. D. Denlinger}
\affiliation{Advanced Light Source, Lawrence Berkeley National Laboratory, Berkeley, CA 94720, USA}
\author{S. Blanco-Canosa}
\affiliation{Donostia International Physics Center (DIPC), San Sebastián, Spain}
\affiliation{IKERBASQUE, Basque Foundation for Science, 48013 Bilbao, Spain}
\author{Jian-Xin Zhu}
\affiliation{Los Alamos National Laboratory, Los Alamos, NM 87545}
\author{E. D. Bauer}
\affiliation{Los Alamos National Laboratory, Los Alamos, NM 87545}
\author{P. F. S. Rosa}
\affiliation{Los Alamos National Laboratory, Los Alamos, NM 87545}

\date{\today}

\maketitle

\newpage

\section{Experimental Methods}
Single crystals of UV$_{6}$Sn$_{6}$ were prepared using the Sn flux technique. U (99.99 \%), V (99.7 \%), and Sn (99.999 \%) pieces in a 1:6:60 ratio were loaded into an alumina crucible and sealed under vacuum in a quartz ampule. The reagents were heated to 1100~$^{\circ}$C, held at 1100~$^{\circ}$C for 12 h, and subsequently slow cooled at 2.5~$^{\circ}$C/h to 800~$^{\circ}$C. Then, the ampule was inverted, and the flux was removed via centrifuge. The resultant hexagonal crystals were up to 3~mm on a side and did not exhibit signs of surface oxidation in air. Scanning electron microscope measurements reveal residual Sn flux on the surface of the crystals, and energy dispersive X-ray measurements yield a stoichiometry of UV$_{6.0(8)}$Sn$_{7.0(7)}$. We attribute the apparent excess of Sn to a combination of residual flux on the surface and the overlap between Sn $L$ (3.44~keV) and U $M$ (3.17~keV) edges.

Magnetic properties of UV$_{6}$Sn$_{6}$ single crystals were collected on a Quantum Design Magnetic Properties Measurement System (MPMS3) with a 7~T magnet. Magnetic susceptibility and isothermal magnetization measurements were collected with fields parallel and perpendicular to the crystallographic $c$ axis.
Additional data up to 16~T was collected using the vibrating sample magnetometer option in a Quantum Design Physical Property Measurement System (PPMS).
Specific heat measurements were obtained in a PPMS with a $^3$He insert capable of reaching 0.35 K and magnetic fields to $B = 9$~T. Measurements were collected between $T = 0.35$ to $300$~K. Nonmagnetic analogue ThV$_{6}$Sn$_{6}$ was measured between $T = 2$ to $300$~K, and the magnetic entropy of UV$_{6}$Sn$_{6}$ was obtained by integrating $C_p/T$ after subtracting off the lattice contributions approximated by ThV$_{6}$Sn$_{6}$. 
Electrical resistivity measurements were obtained in a PPMS with a low-frequency AC resistance bridge. Pt wires were attached to hexagonal crystals in both longitudinal and Hall configurations using silver paint. 
Thermal expansion and magnetostriction measurements were performed using a capacitance dilatometer described in Ref.~27 in the main text.
All thermal expansion and magnetostriction measurements were performed using a slow continuous temperature or magnetic field ramp, respectively.

The crystal structure was determined using a Bruker D8 VENTURE KAPPA single crystal X-ray diffractometer with an I$\mu$S 3.0 microfocus source (Mo K$\alpha$, $\lambda$ = 0.71073 \AA), a HELIOS optics monochromator, and a PHOTON II CPAD detector.
All data were integrated with SAINT V8.41~\cite{SAINT}.
A Multi-Scan absorption correction using SADABS 2016/2 was applied~\cite{Krause}.
Starting models were obtained using intrinsic phasing methods in SHELXT 2018/2 and refined by full-matrix least-squares methods against F2 using SHELXL-2019/2~\cite{Sheldrick1,Sheldrick2}.

ARPES measurements were performed at beamline 4.0.3 (Merlin) of the Advanced Light Source (ALS), Lawrence Berkeley National Laboratory (LBNL). 
Single crystal samples of UV$_{6}$Sn$_{6}$ were attached to sample holders using conductive silver epoxy. Ceramic posts were then glued to the sample surfaces for \textit{in-situ} cleaving. To retrieve cleaved sample pieces, the ceramic posts were additionally glued to the holders with copper wires. The cleaved samples were then heated up to $40$~K for measurements in the paramagnetic state. Spectra were acquired with a Scienta R8000 analyzer. To obtain the $k_{z}$-dependent band dispersion, photon energies were set to the range between 70~eV and 130~eV with a 2~eV step. Photon energies for Fermi surface and valence band measurements are given in each figure.

Neutron powder diffraction was carried out on the high-resolution powder diffractometer for thermal neutrons (HRPT) at PSI~\cite{SINQ}.
For the experiments, 3.709 g of powder were loaded into a vanadium can.
Diffraction data were taken using neutrons of wavelength $2.45\,\mathrm{\AA}$ at temperatures 1.6\,K (foreground) and 100\,K (background). 
Due to uncertainties related to the crystal structure of UV$_6$Sn$_6$, the magnetic structure was determined with refinements on the temperature subtracted scattering intensity (1.6 K - 100 K), in order to isolate magnetic intensity.
The thermal change of lattice parameters leads to a shift of structural Bragg peaks to larger scattering angles at lower temperatures.
In our data, this resulted in lambda-type oversubtractions for scattering angles $2\theta>50$\,deg.
Magnetic structure refinements were therefore restricted to scattering angles $2\theta<50$\,deg.

\newpage

\section{Crystalline Electric Field Analysis}

Our CEF model assumes that the U site in the undistorted lattice has a six-fold rotation symmetry about $c$ and four symmetry-allowed parameters in the CEF Hamiltonian in the Stevens Operator formalism [Refs. 21-22 in the main text]: $\mathcal{H}_{CEF} = B_2^0 O_2^0 +  B_4^0 O_4^0 +  B_6^0 O_6^0 +  B_6^6 O_6^6$, where $O_n^m$ are the CEF Stevens Operators with the quantization axis along $c$ and $B_n^m$ are scalar CEF parameters:

Table~\ref{tab:U3_CEF_params} lists the eigenstates obtained from CEF fits for U$^{3+}$ to the UV$_6$Sn$_6$ susceptibility data.

Table~\ref{tab:EigenvectorsU3_B4} shows the resulting eigenstates from a fit to the U$^{3+}$ ($J=9/2$) models that includes all allowed Stevens parameters.
Importantly, we find that all the fitted models (regardless of how many Stevens parameters were varied) converge to $B_2^0 < 0$ and a ground state doublet dominated by $|\pm 9/2 \rangle$.

\begin{table}[!h]
	\caption{Fitted CEF parameters for $\rm{UV_6Sn_6}$ assuming U$^{3+}$, in units of meV. Non-fitted values are taken from a point-charge model calculation.}
	\begin{ruledtabular}
		\begin{tabular}{c|cccc}
			$B_n^m$ &  Fit $ B_2^0$ & Fit [$ B_2^0$, $ B_4^0$]  & Fit [$ B_2^0$, $ B_4^0$, $ B_6^0$] & Fit all \tabularnewline
			\hline 
			$ B_2^0$ & -6.007$\times 10^{-1}$ & -2.316$\times 10^{-1}$ & -4.465$\times 10^{-1}$ & -2.544$\times 10^{-1}$ \tabularnewline
			$ B_4^0$ & 1.494$\times 10^{-3}$ & -1.119$\times 10^{-2}$ & -4.120$\times 10^{-3}$ & -8.917$\times 10^{-3}$ \tabularnewline
			$ B_6^0$ & -4.639$\times 10^{-5}$ & -4.639$\times 10^{-5}$ & 4.776$\times 10^{-5}$ & -1.484$\times 10^{-4}$ \tabularnewline
			$ B_6^6$ & -6.574$\times 10^{-4}$ & -6.574$\times 10^{-4}$ & -6.574$\times 10^{-4}$ & -3.061$\times 10^{-3}$ \tabularnewline
	\end{tabular}\end{ruledtabular}
	\label{tab:U3_CEF_params}
\end{table}

\begin{table*}[!ht]
	\caption{Eigenvectors and eigenvalues for UV$_6$Sn$_6$ CEF model assuming U$^{3+}$ fitting $B_2^0$, $B_4^0$,  $B_6^0$, and $B_6^6$.}
	\begin{ruledtabular}
		\begin{tabular}{c|cccccccccc}
			E (meV) &$| -\frac{9}{2}\rangle$ & $| -\frac{7}{2}\rangle$ & $| -\frac{5}{2}\rangle$ & $| -\frac{3}{2}\rangle$ & $| -\frac{1}{2}\rangle$ & $| \frac{1}{2}\rangle$ & $| \frac{3}{2}\rangle$ & $| \frac{5}{2}\rangle$ & $| \frac{7}{2}\rangle$ & $| \frac{9}{2}\rangle$ \tabularnewline
			\hline 
			0.000 & -0.9347 & 0.0 & 0.0 & 0.0 & 0.0 & 0.0 & -0.3553 & 0.0 & 0.0 & 0.0 \tabularnewline
			0.000 & 0.0 & 0.0 & 0.0 & 0.3553 & 0.0 & 0.0 & 0.0 & 0.0 & 0.0 & 0.9347 \tabularnewline
			25.814 & 0.0 & -0.5319 & 0.0 & 0.0 & 0.0 & 0.0 & 0.0 & -0.8468 & 0.0 & 0.0 \tabularnewline
			25.814 & 0.0 & 0.0 & -0.8468 & 0.0 & 0.0 & 0.0 & 0.0 & 0.0 & -0.5319 & 0.0 \tabularnewline
			27.328 & 0.0 & 0.0 & 0.0 & 0.0 & 1.0 & 0.0 & 0.0 & 0.0 & 0.0 & 0.0 \tabularnewline
			27.328 & 0.0 & 0.0 & 0.0 & 0.0 & 0.0 & 1.0 & 0.0 & 0.0 & 0.0 & 0.0 \tabularnewline
			30.406 & -0.3553 & 0.0 & 0.0 & 0.0 & 0.0 & 0.0 & 0.9347 & 0.0 & 0.0 & 0.0 \tabularnewline
			30.406 & 0.0 & 0.0 & 0.0 & 0.9347 & 0.0 & 0.0 & 0.0 & 0.0 & 0.0 & -0.3553 \tabularnewline
			60.064 & 0.0 & 0.0 & 0.5319 & 0.0 & 0.0 & 0.0 & 0.0 & 0.0 & -0.8468 & 0.0 \tabularnewline
			60.064 & 0.0 & 0.8468 & 0.0 & 0.0 & 0.0 & 0.0 & 0.0 & -0.5319 & 0.0 & 0.0 \tabularnewline
	\end{tabular}\end{ruledtabular}
	\label{tab:EigenvectorsU3_B4}
\end{table*}

\newpage

\section{Domain physics}

\begin{figure*}[!ht]
	\includegraphics[width=1.0\textwidth]{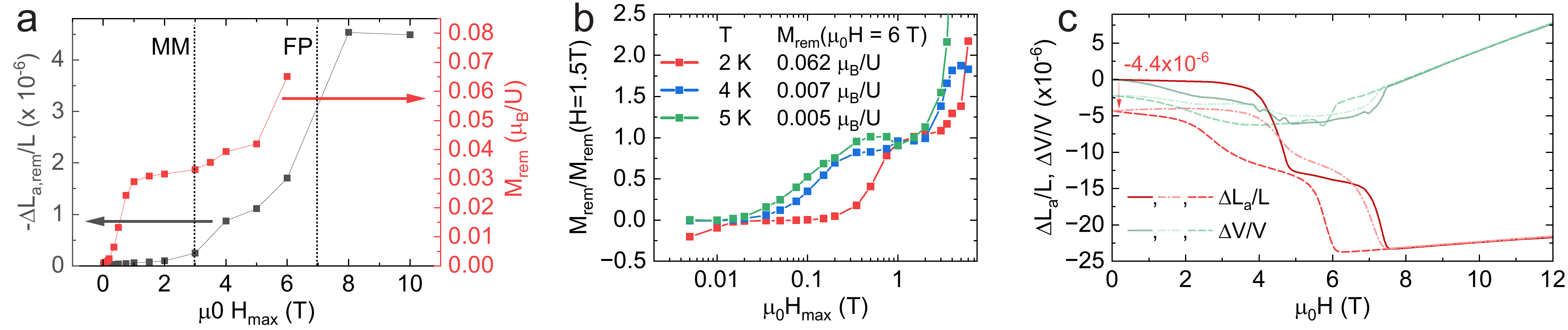}
	\caption{
        (a)~The remanent properties of the sample after cooling in zero field, ramping to the indicated field, ramping back to zero field, and then measuring the indicated quantity at 2~K.
        The left axis (gray symbols) shows the change in the $a$ axis length of the sample.
        The right axis (red symbols) shows the remanent magnetization.
        (b)~The normalized remanent magnetization at different temperatures versus maximum field.
        The legend lists the value of the remanent magnetization at 6~T for the indicated temperatures.
        (c)~Red curves: the change in length along the $a$ axis divided by the length after cooling in zero field.
        Green curves: the change in volume divided by the volume after cooling in zero field.
        The meaning of the solid, dashed-dotted, and dashed lines follows the convention in Fig.~4d in the main text (ZFC, FP Up, and FP Down, respectively).
    }
	\label{fig:domains}
\end{figure*}

To further investigate the unusual remanent magnetization in UV$_6$Sn$_6$, $M_{\text{rem}}$ data were systematically obtained from the following steps: (1) cooling the sample from T~$>$~$T_{\text{N1}}$ to low temperatures in zero field, (2) increasing the \textit{c}-axis field to $H_{\text{max}}$, (3) reducing the field back to zero, and (4) measuring the magnetic moment, shown in Fig.~\ref{fig:domains}a (red symbols, right axis) for $T=$~2~K.
A large increase in $M_{\text{rem}}$ occurs even before the metamagnetic transition, which indicates that the initial change in remanent magnetization is not tied to a phase boundary.
This strongly suggests that $M_{\text{rem}}$ is due to domain reorientation along the \textit{c} axis and that the absence of a net moment in zero-field cooled conditions is due to domain physics rather than a change in the zero-field magnetic structure.

$M_{\text{rem}}$ measurements were also performed at 4~K and 5~K to confirm that the low-field change is not due to flux pinning from a potential V$_3$Sn inclusion, whose superconducting transition temperature is expected at T$_c$ = 3.6~K.
Fig.~\ref{fig:domains}b shows $M_{\text{rem}}$ at 4~K and 5~K normalized to the value at 1.5~T, which corresponds to the plateau region common to all three temperatures.
The field at which $M_{\text{rem}}$ decreases with temperature, which is consistent with domain behavior because the domain pinning is expected to become weaker as the temperature is increased.

Although the lack of a net moment in zero field cooled conditions can be attributed to magnetic domains, there is also an indication that the zero-field structure may be different after field sweeps that go above $H_{p}$.
Figure~\ref{fig:domains}c shows that similarly to the $c$ axis, there is a change in length along the $a$ axis after ramping the field to the fully-polarized state and back to zero.
This length change was measured as a function of maximum applied field using a similar procedure as described above for remanent magnetization, and the results are shown in Fig.~\ref{fig:domains}a (gray symbols, left axis).
Unlike $M_{\text{rem}}$, the ``remanent'' $\Delta L_{a,rem}$ does not show an appreciable increase in the low-field region below $H_{p}$.
There is instead a more substantial change in $\Delta L_{a,rem}$ just above $H_{p}$ that then goes through another step-length increase at the phase boundary to the fully-polarized state.
Although not measured into the fully-polarized state due to limitations in the maximum magnetic field, the remanent magnetization seems to be following these trends in the length change.
This result shows that there is an initial domain reconfiguration at low fields that is more pronounced  in the remanent magnetization, but that additional changes occur at higher fields that affect both the remanent magnetization and the length of the sample.
Neutron scattering experiments with different field histories would be useful to understand these changes.

\newpage

\section{Additional electronic transport and thermodynamic data}

\begin{figure*}[!ht]
	\includegraphics[width=0.5\textwidth]{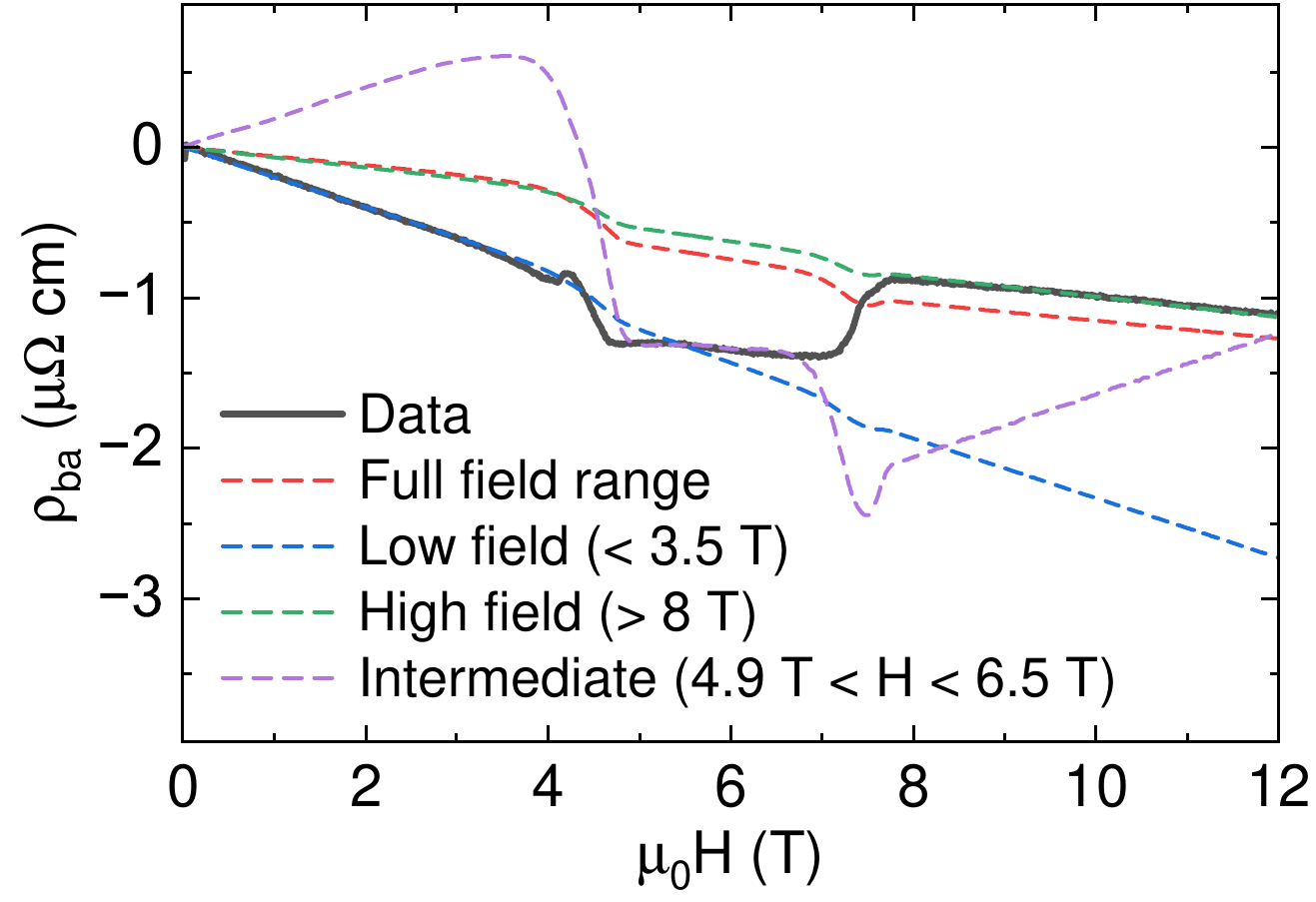}
	\caption{
        Attempts to fit the Hall resistivity to an ordinary and anomalous contribution: $\rho_{ba} = R_OH + R_A\rho^2_{aa}(H)M(H)$.
        The different dashed lines correspond to fitting the data over the indicated field ranges.
    }
	\label{fig:Hall}

\end{figure*}

To look for evidence of a topological contribution to the Hall resistivity, the Hall resistivity was fit using the following functional forms: (1) $\rho_{ba} = R_OH + R_A\rho_{aa,H=0}M(H)$, (2) $\rho_{ba} = R_OH + R_A\rho_{aa}(H)M(H)$, and (3) $\rho_{ba} = R_OH + R_A\rho^2_{aa}(H)M(H)$.
Here $R_O$ and $R_A$ are fit parameters corresponding to the ordinary and anomalous contributions to the Hall resistivity, respectively.
It was not possible to fit the Hall resistivity to the entire measured field range for any of the forms.
This likely indicates that $R_O$ and $R_A$ change as a function of field due to an underlying change in the Fermi surface.
As an example the fits for $\rho_{ba} = R_OH + R_A\rho^2_{aa}(H)M(H)$ are shown in Fig.~\ref{fig:Hall}.

\begin{figure*}[!ht]
	\includegraphics[width=\textwidth]{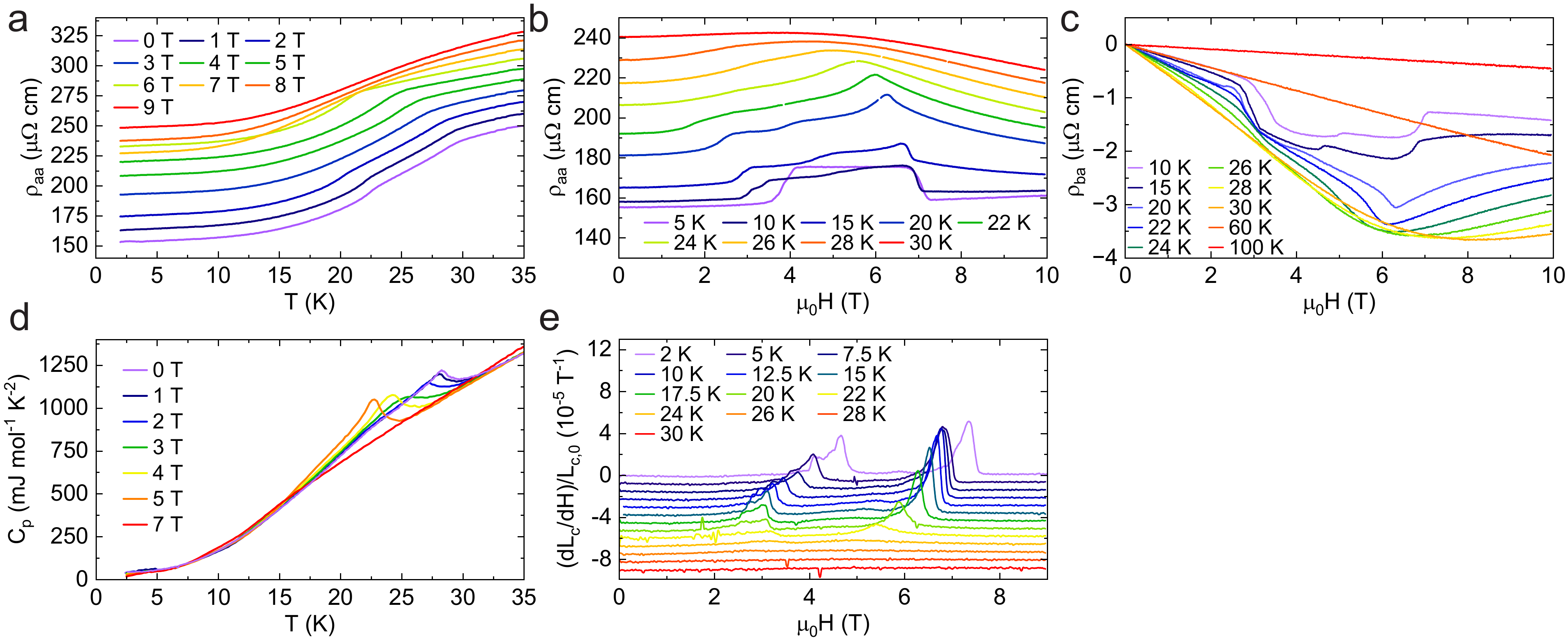}
	\caption{
        (a)~In-plane resistivity versus temperature at different magnetic fields. Below 3~T, two clear transitions are observed.
        (b)~In-plane resistivity versus field at different temperatures.
        (c)~In-plane hall resistivity versus field at different temperatures.
        For (b) and (c) the data was symmetrized or anti-symmetrized, respectively, using measurements in positive and negative fields while increasing the magnitude of the field.
        At intermediate temperatures, additional features are observed in the resistivity.
        These were not associated with a phase boundary due to the lack of similar features in thermodynamic probes.
        (d)~Specific heat versus temperature at different magnetic fields.
        (e)~The derivative of the magnetostriction with respect to field versus field.
    }
	\label{fig:field}

\end{figure*}

Figure~\ref{fig:field} provides supplementary electronic transport and thermodynamic data used to derive the phase diagram shown in the main text.

~

\begin{figure*}
    \includegraphics[width=0.5\textwidth]{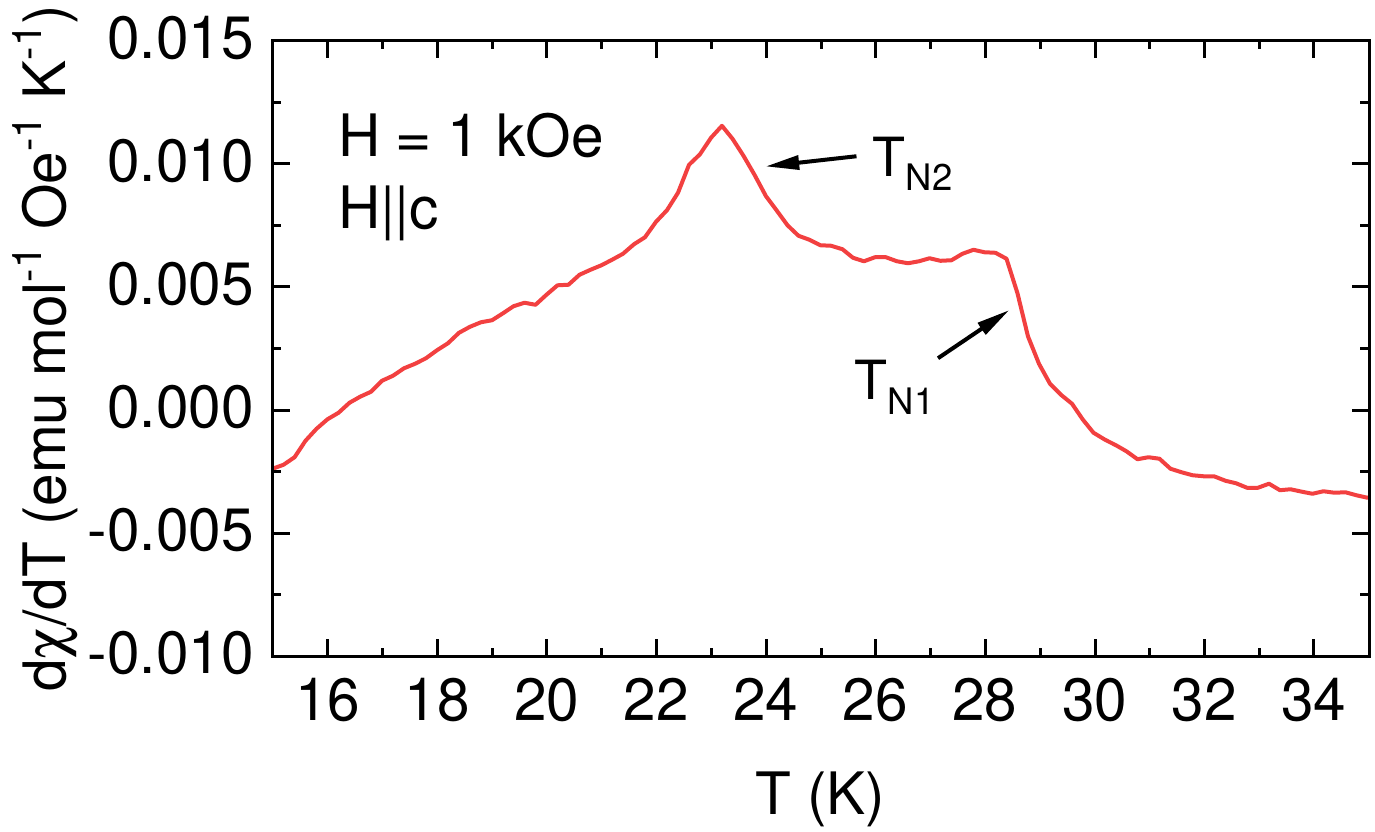}
	\caption{
            The derivative of the susceptibility versus temperature for field applied along the $c$ axis.
            There are clear features at T$_{N1}$ and T$_{N2}$ as indicated.
        }

	\label{fig:chideriv}

\end{figure*}

~

\newpage

\section{Crystal structure determination from x-ray diffraction}

Here we report on the crystallographic data and atomic position as determined from laboratory single-crystal x-ray diffraction.
This report and the CIF file were generated using FinalCif~\cite{FinalCIF}.
The average structure is best modeled in P2/m with a q vector of (-0.144a*, 0.5b*, 0.077c*).
Although the partial occupancy and modulation along b* means that the parent compound is more accurately described as U$_{0.5}$V$_{3}$Sn$_{3}$, we refer to it in the main text as UV$_{6}$Sn$_{6}$ for simplicity.
Table~\ref{tab:crystallographic-data} provides the crystallographic data and refinement parameters and table~\ref{tab:atomic-pos} presents the atomic positions. 

\begin{table}[!h]
	\caption{Crystallographic data and refinement parameters.
    }
		\begin{tabular}{ |c|c| }
            \hline
			Space group (number) & P2/m (10) \\ \hline
            a [\AA] & 5.5197(3) \\ \hline
            b [\AA] & 4.5857(2) \\ \hline
            c [\AA] & 5.5251(3) \\ \hline
            $\alpha{}$ [$^{\circ}$] & 90 \\ \hline
            $\beta{}$ [$^{\circ}$] & 119.965(2) \\ \hline
            $\gamma{}$ [$^{\circ}$] & 90 \\ \hline
            Volume[\AA$^3$] & 121.156(11) \\ \hline
            Z & 1 \\ \hline
            $\mu$ [mm$^{-1}$] & 37.208 \\ \hline
            Crystal size [mm$^3$] & 0.039 $\times$ 0.052 $\times$ 0.071 \\ \hline
            $2\theta$ range [$^{\circ}$] & 8.52 to 60.68 (0.70 \AA) \\ \hline
            Reflections collected & 5173 \\ \hline
            Unique reflections & 411 \\ \hline
            Goodness-of-fit on $F^2$ & 1.013 \\ \hline
            Final R indexes [$l\ge{}2\sigma{}(l)$] & \begin{tabular}{@{}c@{}}R$_1$ = 0.0227 \\ wR$_2$ = 0.0524\end{tabular} \\ \hline
            Largest peak/hole [e\AA$^{-3}$] & 1.44/-2.34 \\ \hline
	\end{tabular}
	\label{tab:crystallographic-data}
\end{table}

\begin{table}[!h]
	\caption{Atomic positions.
    }
		\begin{tabular}{ |m{2cm}|m{2cm}|m{2cm}|m{2cm}|m{2cm}|m{2cm}| }
            \hline
			\textbf{Atom} & \textbf{x} & \textbf{y} & \textbf{z} & \textbf{Occupancy} & \textbf{U$_{eq}$} \\ \hline
            U1 & 0 & 0 & 0.5 & 0.5 & 0.00749(17) \\ \hline
            Sn1 & 0 & 0.32317(11) & 0.5 & 0.5 & 0.00635(19) \\ \hline
            Sn2 & 0.66668(4) & 0 & 0.83330(4) & 1 & 0.00590(17) \\ \hline
            V1 & 0.5 & 0.5 & 0.5 & 1 & 0.0053(2) \\ \hline
            V2 & 0.5 & 0.5 & 0 & 1 & 0.0055(2) \\ \hline
            V3 & 0 & 0.5 & 0 & 1 & 0.0055(2) \\ \hline
	\end{tabular}
	\label{tab:atomic-pos}
\end{table}

\newpage

\section{Magnetic neutron powder diffraction of UV$_6$Sn$_6$}
For the group theoretical symmetry analysis of the magnetic structure refinement shown in the main text, we considered by way of simplification a hexagonal lattice of uranium atoms as described in parent space group $P6/mmm$.
Representational analysis with wave-vector $(0,0,0.5)$ for the Uranium atom at (0,0,0), carried out with the software SARAh \cite{2000_Wills_PhysicaBCondensedMatter}, provides two irreducible representations (IRs), namely $\Gamma_3$, which has one basis vector along the $c$-axis, and $\Gamma_9$, which has two basis vectors in the basal hexagonal plane.

The magnetic scattering, shown in Fig. 3 of the main text, was indexed within a hexagonal lattice of parameters $a=5.52\,\mathrm{\AA}$ and $c=9.17\,\mathrm{\AA}$ (conventional unit cell parameters inferred from structural Bragg peaks recorded at 1.6 K) and a magnetic propagation vector $\bm{k}=(0,0,\frac{1}{2})$.
The magnetic diffraction pattern suggests that the magnetic structure factor is suppressed for momentum transfers lying on the reciprocal $L$-axis, whereas for reciprocal-space locations with a finite $H$- or $K$-component, the structure factor is finite.
Together, this points towards antiferromagnetic order with moments parallel to the $c$-axis.

To illustrate this quantitatively, magnetic structure refinements were carried out with the software Fullprof \cite{rodriguez-carvajalRecentAdvancesMagnetic1993}.
We optimized the free parameters allowed in the respective irreducible representation (1 basis vector for $\Gamma_3$ and 2 basis vectors for $\Gamma_9$).
The parameters that describe the widths of the Bragg peaks were inferred from structural Bragg peaks recorded in the paramagnetic state and held constant.

As the magnetic ion, we considered U$^{3+}$, and approximated the form factor with the expansion:

\begin{align}
    f= \left\langle j_0 \right\rangle + c_2\cdot\left\langle j_2 \right\rangle \, .
 \end{align}
 
For the radial integrals of spherical Bessel functions, $\left\langle j_n \right\rangle$, we took the values tabulated in Ref. \cite{1976_Freeman_PhysRevB}.
For the constant, $c_2$, we took the value obtained for the Hunds-rule Russell-Saunders ground state given by $c_2=1.75$ (see also Ref. \cite{1976_Freeman_PhysRevB}).
Good agreement for $\Gamma_3$ in our refinements is indicated by $\chi^2=1.41$ and magnetic $R$-value 35.87. For $\Gamma_9$, we instead obtained $\chi^2=1.87$ and $R$-value 91.94. 

\subsection{Crystal structure refinements}
An estimate for the size of the ordered magnetic moment in UV$_6$Sn$_6$ was inferred fitting the nuclear Bragg scattering (taken in the paramagnetic state at 100 K with wavelength $2.45\,\mathrm{\AA}$) with the crystal-structure model listed in Table \ref{tab:CrystalStructure}.
The monoclinic structure has one uranium atom per unit cell with 50 percent site-occupancy.
In the refinement, we optimized an overall scale-parameter, the cell parameters of the monoclinic unit cell, the parameters describing the widths of the Bragg peaks, and the $y$- and $z$-coordinate of the atom Sn1 and Sn2, respectively.
The background was fitted with a polynomial function of 6th order.

\begin{figure*}[!h]
    \centering
    \includegraphics[width=0.6\columnwidth]{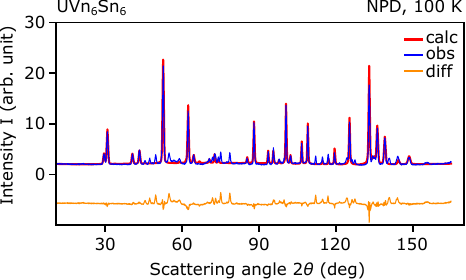}
    \caption{Neutron powder diffraction (NPD) data of UV$_6$Sn$_6$ taken in the paramagnetic state. The observed intensity (blue solid line) was fit with a monoclinic structure associated with space-group $P2/m$. The red line corresponds to the calculated diffraction pattern. The orange line corresponds to the difference of observed and calculated intensity and was shifted by a y-offset of -5.6.
    }\label{fig:FigureSuppX}
\end{figure*}

In addition to the phase of interest, we considered the impurity phases V$_3$Sn (22 percent), USn$_3$ (1.5 percent), and Sn (0.1 percent).
Fig. \ref{fig:FigureSuppX} (a) shows the comparison of the fit ($\chi^2=43.15$) with the recorded data.
Many nuclear scattering peaks are not accounted for, even with the impurity phases.
Given the ambiguity of the UV$_6$Sn$_6$ crystal structure, such a mismatch is not unexpected.
Nevertheless, this rough nuclear refinement captures the intensities of the main nuclear Bragg peaks, from which we infer the resolution width and the overall scale factor. 
Magnetic structure refinement in monoclinic symmetry with the scale factor from the structural refinement results in an ordered moment of $1.66(48)\,\mu_{\mathrm{B}}$ per uranium atom. 

\begin{table}[b]
\caption{Crystal-structure model of UV$_6$Sn$_6$ that was considered for the refinement. The table lists the atomic types, the label of the Wyckoff position, the respective coordinates, the occupancy as used in Fullprof, and the relative occupancy (measured with respect to the full occupancy of the respective Wyckoff position).
The structure is monoclinic with space group $P2/m$, lattice parameters $a=5.5200(5)\,\mathrm{\AA}$, $b=4.5812(4)\,\mathrm{\AA}$, and $c=5.5167(7)\,\mathrm{\AA}$, and angular parameters $\alpha=90$ deg, $\beta=120.039(7)$ deg, and $\gamma=90$ deg.\label{tab:CrystalStructure}}
\begin{ruledtabular}
\begin{tabular}{cccccc}
Atom & Wyckoff Position &Coordinates&Occupancy&Relative Occupancy\\
\hline
U&c&(0, 0, 1/2)&0.125&0.5\\
Sn1&k&(0, 0.3186(10), 1/2)&0.25&0.5\\
Sn2&m&(2/3, 1, 0.8213(13))&0.5&1\\
V1&h&(1/2, 1/2, 1/2)&0.25&1\\
V2&e&(1/2, 1/2, 1)&0.25&1\\
V3&b&(1, 1/2, 1)&0.25&1\\
\end{tabular}
\end{ruledtabular}
\end{table}

~

~

~

~

~

\newpage
\section{Additional ARPES data}

\begin{figure*}[!ht]
	\includegraphics[width=0.75\textwidth]{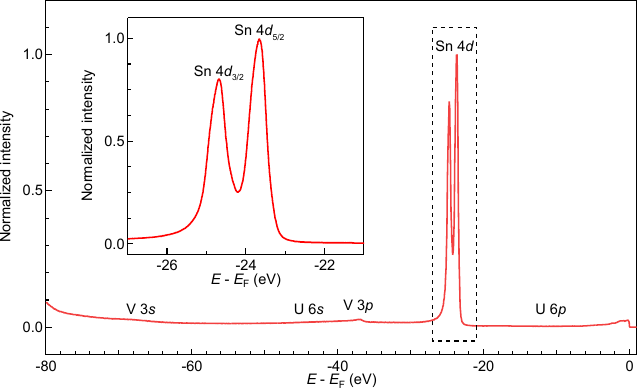}
	\caption{Core-level spectrum of UV$_{6}$Sn$_{6}$, normalized to the maximum intensity. The inset provides a zoomed-in view of the region within the dashed box. The spectrum is dominated by Sn $4d$ peaks without additional shoulders, indicating a Sn$_{2}$ termination similar to that of ScV$_{6}$Sn$_{6}$ [see Ref. 33 in the main text].}
	\label{fig:ARPES_core}
\end{figure*}

Fig.~\ref{fig:ARPES_core} shows the core-level spectrum measured with 102~eV photons. The intensity of the spectrum is dominated by Sn $4d$ peaks, but other small peaks, such as V $3s$ and $3p$, are also visible. The inset presents a zoomed-in image of the Sn $4d$ core peaks in the black dashed box, where two peaks correspond to the Sn $4d_{3/2}$ and $4d_{5/2}$ states without any additional peaks. In ScV$_{6}$Sn$_{6}$, shoulders in Sn core peaks were reported for terminations other than the Sn$_{2}$ termination [see Ref. 33 in the main text], thus we conclude that the observed electronic structures of UV$_{6}$Sn$_{6}$ originate from the Sn$_{2}$ termination.

\begin{figure*}[!ht]
	\includegraphics[width=0.8\textwidth]{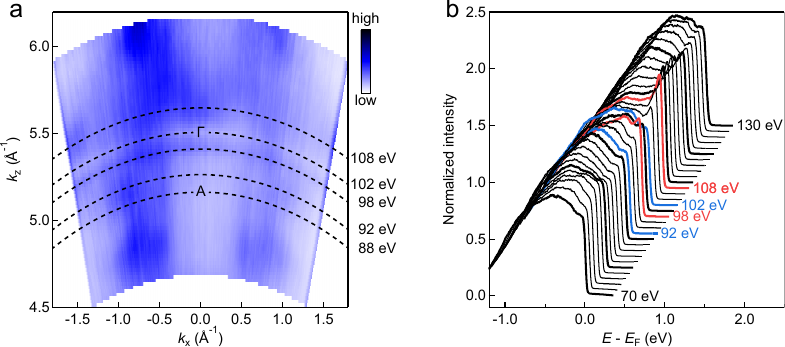}
	\caption{Photon-energy dependent ARPES results. (a) $k_{x}-k_{z}$ plane Fermi surface mapped using photon energy dependence measurements. The inner potential $V_{0} = 18$~eV is used. Black dashed lines indicate constant photon energy cuts at 88, 92, 98, 102, and 108 eV. The $\Gamma$ and $A$ points are accessible with 102 eV and 88 eV photons, respectively. (b) $k_{x}$-integrated energy distribution curves (EDCs) from 70 eV to 130 eV with 2 eV step. EDCs at every 10 eV highlighted in bold. Compared to off-resonance conditions (blue, 92, 102 eV), on-resonance EDCs (red, 98, 108 eV) exhibit small peaks near the Fermi level due to U $5f$ weight. }
	\label{fig:ARPES_kz}

\end{figure*}

Figure~\ref{fig:ARPES_kz} shows photon energy-dependent ARPES data. Figure~\ref{fig:ARPES_kz}a presents the $k_{x}-k_{z}$ plane Fermi surface of UV$_{6}$Sn$_{6}$, measured using photons in the 70~eV to 130~eV energy range. The inner potential $V_{0}$ is set to 18 eV, and 102 eV and 88 eV correspond to the $\Gamma$ and $A$ points, respectively. Photon energies of 98~eV and 108~eV represent on-resonance conditions for U $f$ electrons and lead to enhanced ARPES intensity near the Fermi level compared to off-resonance conditions at 92~eV and 102~eV. This effect is more clearly observed in the $k_{x}$-integrated energy distribution curves (EDCs) from 70~eV to 130~eV, shown in Figure~\ref{fig:ARPES_kz}b. In addition to the 108~eV resonance presented in the main text, the 98 eV EDC also exhibits a slight enhancement of ARPES intensity near the Fermi level compared to off-resonance conditions at 92 eV and 102 eV.

\begin{figure*}[!ht]
	\includegraphics[width=0.5\textwidth]{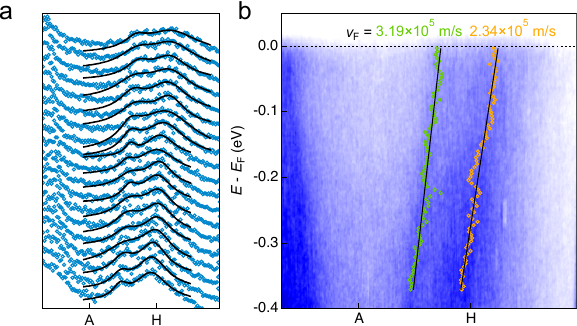}
	\caption{Momentum distribution curve (MDC) fit for extracting $v_{F}$. (a) Stacked MDCs (blue markers) with corresponding fit lines (black lines). (b) MDC peak positions for the A-H inner (green) and outer (orange) pockets. Black solid lines are linear fits of the peak positions and used to determine $v_{F}$ for both pockets.}
	\label{fig:ARPES_MDC}
    \end{figure*}

Figure~\ref{fig:ARPES_MDC} shows the momentum distribution curve (MDC) fit and extraction of $v_{F}$ for two electron bands in the dispersion 
along the A-H direction. MDCs of these bands were fitted with two Lorentzian peaks convolved with a Gaussian function, as shown in 
Fig.~\ref{fig:ARPES_MDC}a, and the extracted peak positions are plotted in Fig.~\ref{fig:ARPES_MDC}b. To determine $v_{F}$, these peak positions 
were fitted with linear functions, represented by the black solid lines in Fig.~\ref{fig:ARPES_MDC}b. From the slopes of these fits, the Fermi 
velocities are calculated using $v_{F}=\frac{1}{\hbar}\frac{dE}{dk}$, yielding $3.19\times10^5$ m/s for the inner pocket (green) and 
$2.34\times10^5$ m/s for the outer pocket (orange).

\vspace{50 cm}
~
\bibliography{lib}